\documentclass[aps,superscriptaddress,twocolumn,twoside,floatfix,pra,nofootinbib,a4paper]{revtex4-2}

\pdfoutput = 1

\usepackage{times}
\usepackage{epsfig}
\usepackage{amsfonts}
\usepackage{amsmath}
\usepackage{amssymb}
\usepackage{amsthm}
\usepackage{color}
\usepackage{multirow}
\usepackage[normalem]{ulem}
\newcommand{\stkout}[1]{\ifmmode\text{\sout{\ensuremath{#1}}}\else\sout{#1}\fi}
\usepackage{latexsym}
\usepackage{mathrsfs}
\usepackage{natbib}
\usepackage{verbatim}
\usepackage[T1]{fontenc}
\usepackage{subfig}
\usepackage{inputenc}
\usepackage{floatrow}
\usepackage{natbib}

\usepackage{floatrow}
\usepackage{array,booktabs}

\usepackage{graphicx}
\usepackage{xcolor}

\usepackage[colorlinks=true,linkcolor=blue,citecolor=magenta,urlcolor=blue]{hyperref}

\DeclareMathOperator{\Tr}{tr}

\newcommand{\ket}[1]{|#1\rangle}

\newcommand{\bracket}[3]{\langle#1|#2|#3\rangle}
\newcommand{\ketbra}[2]{|#1\rangle\langle#2|}

\begin{document}

%%%%%%%%%%%%%%%%%%%%%%%%%%%%%%%%%%%%%%%%%%%%%%%%%%%%%%%%%%%%%%%%%%%
\title{Entanglement-assisted quantum communication with simple measurements}

\author{Amélie Piveteau}
\affiliation{Department of Physics, Stockholm University, S-10691 Stockholm, Sweden}

\author{Jef Pauwels}
\affiliation{Laboratoire d'Information Quantique, CP 225, Universit\'e libre de Bruxelles (ULB),\\ Av. F. D. Roosevelt 50, 1050 Bruxelles, Belgium}

\author{Emil H\aa kansson}
\affiliation{Department of Physics, Stockholm University, S-10691 Stockholm, Sweden}
\affiliation{Hitachi Energy Research, Forskargränd 7, 72219 V\"aster\aa s, Sweden}

\author{Sadiq Muhammad}
\affiliation{Department of Physics, Stockholm University, S-10691 Stockholm, Sweden} 
\affiliation{Department of Applied Physics, Royal Institute of Technology (KTH), Stockholm 106 91, Sweden.}

\author{Mohamed Bourennane}\thanks{Corresponding author: boure@fysik.su.se}
\affiliation{Department of Physics, Stockholm University, S-10691 Stockholm, Sweden}

\author{Armin Tavakoli}
\affiliation{Institute for Quantum Optics and Quantum Information - IQOQI Vienna, Austrian Academy of Sciences, Boltzmanngasse 3, 1090 Vienna, Austria}
\affiliation{Atominstitut,  Technische  Universit{\"a}t  Wien, Stadionallee 2, 1020  Vienna,  Austria}

\begin{abstract}
Dense coding is the seminal example of how entanglement can boost qubit communication, from sending one bit to sending two bits. This is made possible by projecting separate particles onto a maximally entangled basis. We investigate more general communication tasks, in both theory and experiment, and show that simpler measurements enable strong and sometimes even optimal entanglement-assisted qubit communication protocols. Using only partial Bell state analysers for two qubits, we demonstrate quantum correlations that cannot be simulated with two bits of classical communication. Then, we show that there exists an established and operationally meaningful task for which product measurements are sufficient for the strongest possible quantum predictions based on a maximally entangled two-qubit state. Our results reveal that there are scenarios in which the power of entanglement in enhancing quantum communication can be harvested in simple and scalable optical experiments.
\end{abstract}

\date{September 22, 2022}
\maketitle

\section*{Introduction}

Entanglement and quantum communication are both paradigmatic resources for quantum information science and crucial for understanding the nonclassical nature of quantum theory. The former has been studied for decades in Bell-type experiments \cite{Hensen2015, Zeilinger2015, Shalm2015, Rosenfeld2017}, where communication between the parties is not allowed. The latter has, in more recent years, been extensively studied in prepare-and-measure experiments, where shared entanglement is absent \cite{Gallego2010, Ahrens2012, Hendrych2012}. It is therefore natural to investigate the most general scenario, featuring both entanglement and quantum communication. 

\begin{figure}[t!]
\centering
\includegraphics[width=0.9\columnwidth]{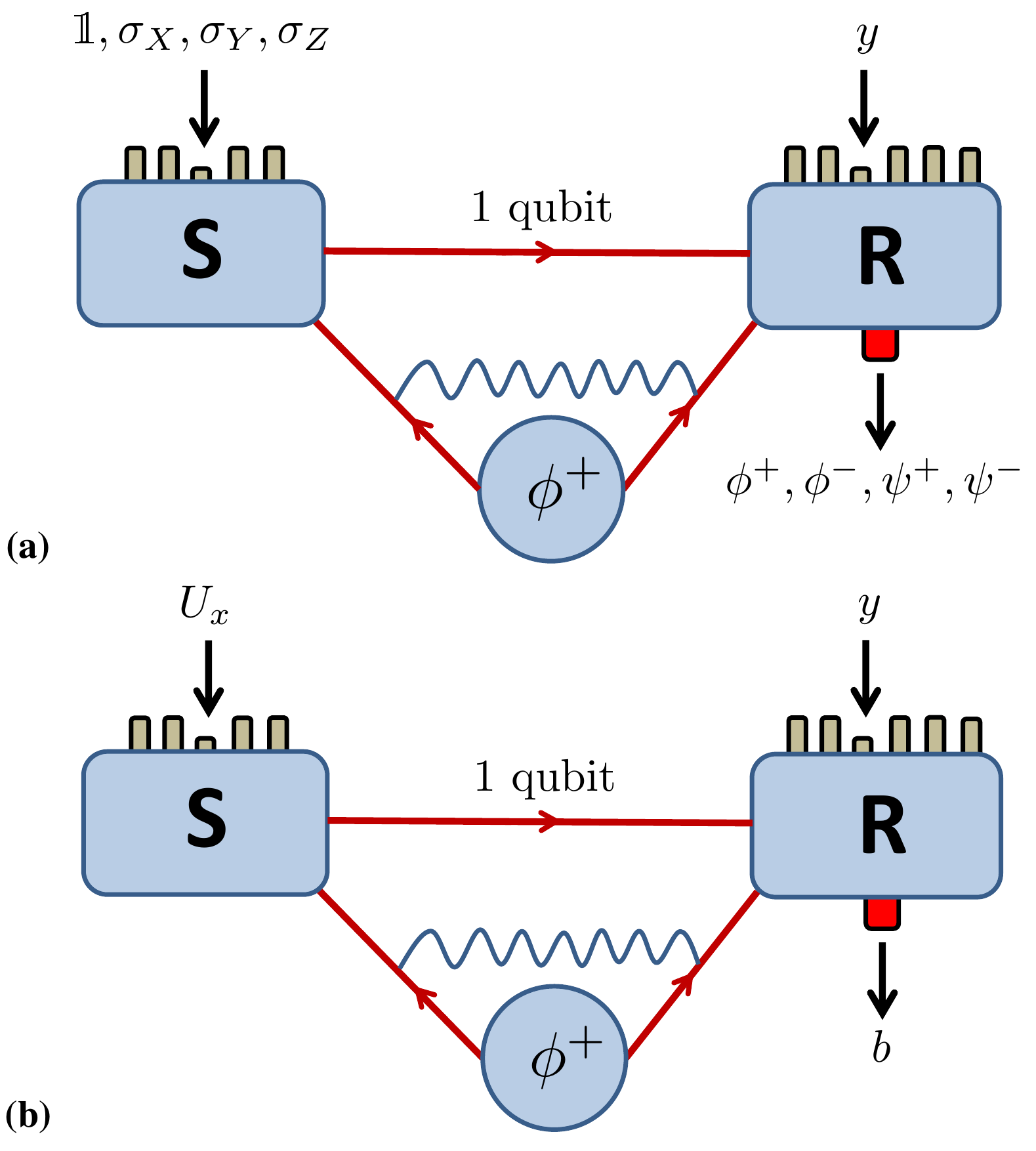}
\caption{(a) \emph{Dense coding scenario}. The sender (S) and receiver (R) share an EPR pair $\ket{\phi^+}$. The sender selects one of four Pauli unitaries $\{\openone,\sigma_X,\sigma_Y,\sigma_Z\}$, applies it to her qubit and relays it to the receiver who performs a measurement of both qubits in the basis of the four Bell states $\ket{\phi^\pm}=\frac{1}{\sqrt{2}}(\ket{00}\pm\ket{11})$ and $\ket{\psi^\pm}=\frac{1}{\sqrt{2}}(\ket{01}\pm\ket{10})$. From the outcome, the sender's two bit input can be recovered. (b) \emph{Generic sender-receiver scenario.} The parties again share an EPR pair and are allowed to communicate a qubit. The sender can now select between any number of arbitrary unitary operations $U_x$ and  the receiver can select between any number of arbitrary quantum measurements. \label{Fig1}}
\end{figure}

Dense coding  is a striking illustration of the power of entanglement-assisted quantum communication \cite{Bennett1992} (Fig.~\ref{Fig1}a). By sharing an Einstein-Podolsky-Rosen (EPR) pair, dense coding allows one to transmit two bits of classical information while sending only one qubit \cite{Bennett1992}. In contrast, a qubit alone can never carry more than one bit of information \cite{Holevo1973}. No entanglement-assisted protocol based on sending a qubit can transmit more than two bits \cite{Bennett1999}. Crucially, in addition to having an EPR pair, dense coding also requires the ability to jointly measure both shares in a basis of four maximally entangled two-qubit states; a so-called Bell basis measurement. For such a  task, separable measurements do not offer any quantitative advantage over standard classical communication, regardless of the type of shared entangled state \cite{Pauwels2022}.

However, in contrast to some other platforms for dense coding \cite{Schaetz2004,Fang2000}, optical systems do not allow a straightforward implementation of a Bell basis measurement on separate photonic carriers. While optics is a particularly natural platform for quantum communication, such an implementation is impossible with linear optics \cite{Luetkenhaus1999, Vaidman1999} unless one employs auxiliary degrees of freedom \cite{Kwiat1998}. Nevertheless, two-qubit optical demonstrations of dense coding have been performed, for example by implementing deterministic partial Bell basis measurements  that can in principle harvest at most $\log_2 3$ bits \cite{Mattle1996}, or by encoding in continuous variables \cite{Braunstein2000, Li2002}, or by means of hyperentanglement which introduces additional photonic qubits \cite{Schuck2006, Barbieri2007, Barreiro2008, Williams2017}. Going beyond qubit systems, partial Bell basis measurements have recently been implemented on entanglement-assisted  systems of dimension four to beat the two bit communication limit  \cite{Hu2018}. Nevertheless, deterministic implementations of sophisticated entangled measurements, in particular without auxiliary qubits, is difficult unless they are considerably restricted. Moreover, scaling a Bell basis measurement beyond the few lowest dimensions is an outstanding challenge.

Here we go beyond the dense coding task and consider more general communication tasks \cite{Wiesner1983, Ambainis1999} implemented with quantum messages assisted by entanglement. In such scenarios, when both entanglement and quantum communication are available, little is known about the predictions of quantum theory \cite{Tavakoli2021a, Pauwels2022a}. Here, we focus on the most elementary resources for such settings, namely a qubit message and a shared EPR state. We find that there exists correlation scenarios in which simple measurements can give rise to quantum correlations that cannot be simulated with two bits of classical communication, i.e.~they cannot be reproduced with an ideal dense coding protocol. To this end, we first introduce a correlation task for which a standard partial Bell state analyser \cite{Weinfurter1994, Braunstein1995} creates  quantum correlations, that cannot be simulated with two bits of communication. Then, we go further and consider a well-established communication task, known as a Random Access Code, and show that product measurements are sufficient not only to elude classical models based on two bits of communication, but  even to achieve the strongest predictions allowed by quantum theory for a two-qubit system. Thus, there exists natural communication tasks that can be implemented optimally by means of a quantum channel assisted by entanglement without the need for interference between the two photonic carriers in the measurement apparatus.

\section*{Results}
Consider a generic communication task in which the sender selects a classical input $x$ and encodes it into a message that is sent to the receiver. The receiver selects a question, labeled $y$, to which he produces an answer labeled $b$ (Fig.~\ref{Fig1}b). After many runs, they obtain probabilities $p(b|x,y)$. The parties pre-share the state $\ket{\phi^+}=\frac{1}{\sqrt{2}}\left[\ket{00}+\ket{11}\right]$ and the message consists of a single qubit, which is encoded via a local unitary $U_x$ on the sender's share. Once the receiver holds both shares, he performs the measurement $\{E_{b|y}\}$. The probabilities are given by the Born rule,
\begin{equation}
p(b|x,y)=\Tr\left(E_{b|y}(U_x\otimes\openone)\phi^+(U_x^\dagger\otimes\openone)\right).
\end{equation}	
Via dense coding, any $p(b|x,y)$, where $x$ takes at most four values, can be generated in the experiment, regardless of the number of questions the receiver asks. Therefore, to find correlations that go beyond dense coding, one needs at least five values of $x$. We use the ability to beat classical communication models based on two bits as a basic benchmark for quantum protocols.

\subsection{Stochastic dense coding with Bell basis measurements}
First, we show that there exists a natural information-theoretic task whose performance can be enhanced beyond what is possible with two bits. Consider a Random Access Code (RAC) \cite{Ambainis1999}: the sender holds $x=x_1x_2\in\{1,2,3,4\}^2$, and the receiver privately and uniformly selects $y\in\{1,2\}$ with the aim of recovering $x_y$. This is a stochastic dense coding task, with average success rate 
\begin{equation}
\mathcal{R}=\frac{1}{32}\sum_{x,y}p(b=x_y|x,y).
\end{equation}
Via dense coding, the receiver can, e.g., always recover $x_1$ but is then forced to guess the value of $x_2$, yielding  $\mathcal{R}=\frac{5}{8}$. In fact, no better two-bit strategy is possible (see Methods). Nevertheless, this bound can be exceeded using the same quantum resources. Let the receiver measure the bases 
\begin{align}
& \ket{E_{b|1}}=\openone\otimes \sigma_X^{b_1}\sigma_Z^{b_2}\ket{\phi^+}\\
& \ket{E_{b|2}}=\openone \otimes R \ket{E_{b|1}},
\end{align}
where $R=\frac{1-i}{2\sqrt{2}}\openone+\frac{1+i}{2\sqrt{2}}(\sigma_X+\sigma_Y+\sigma_Z)$  and $b=b_1b_2\in\{0,1\}^2$. Given these measurements, the success rate is bounded by
\begin{align}\nonumber
\mathcal{R}&\leq \frac{1}{32} \sum_x \max_{\{\ket{\psi_x}\}} \bracket{\psi_x}{ E_{x_1|1}+E_{x_2|2}}{\psi_x} \\
&=\frac{1}{32}\sum_x \lambda_\text{max}(E_{x_1|1}+E_{x_2|2})=\frac{3}{4},
\end{align}
where $\lambda_\text{max}$ is the eigenvalue with the largest magnitude. This bound is reachable in our scenario because the eigenvector corresponding to $\lambda_\text{max}$ for each $x$ (the optimal two-qubit state) is maximally entangled, and hence realisable via a local unitary on $\ket{\phi^+}$. No better quantum protocol exists because the best protocol based on sending an unassisted four-dimensional quanta is known to achieve $\mathcal{R}=\frac{3}{4}$  \cite{Tavakoli2015}, which must strictly bound the protocols of our interest from above.  We refer to to Supplementary Note 1 for a more detailed discussion of the Random Access Code, including the relation between its success probability and the classical capacity of (entanglement-assisted) communication channels.

However, the above advantage is based on performing a pair of (rotated) Bell basis measurements, i.e.~measurements similar to that used in dense coding. These lack both simple implementation and scalability in dimension. Moreover, even though essential for dense coding, it may not be that such sophisticated measurements are necessary for more general entanglement-assisted quantum communications.	We therefore proceed to investigate the usefulness of considerably more elementary measurements.

\begin{figure*}[t!]
  \includegraphics[width=0.8\textwidth]{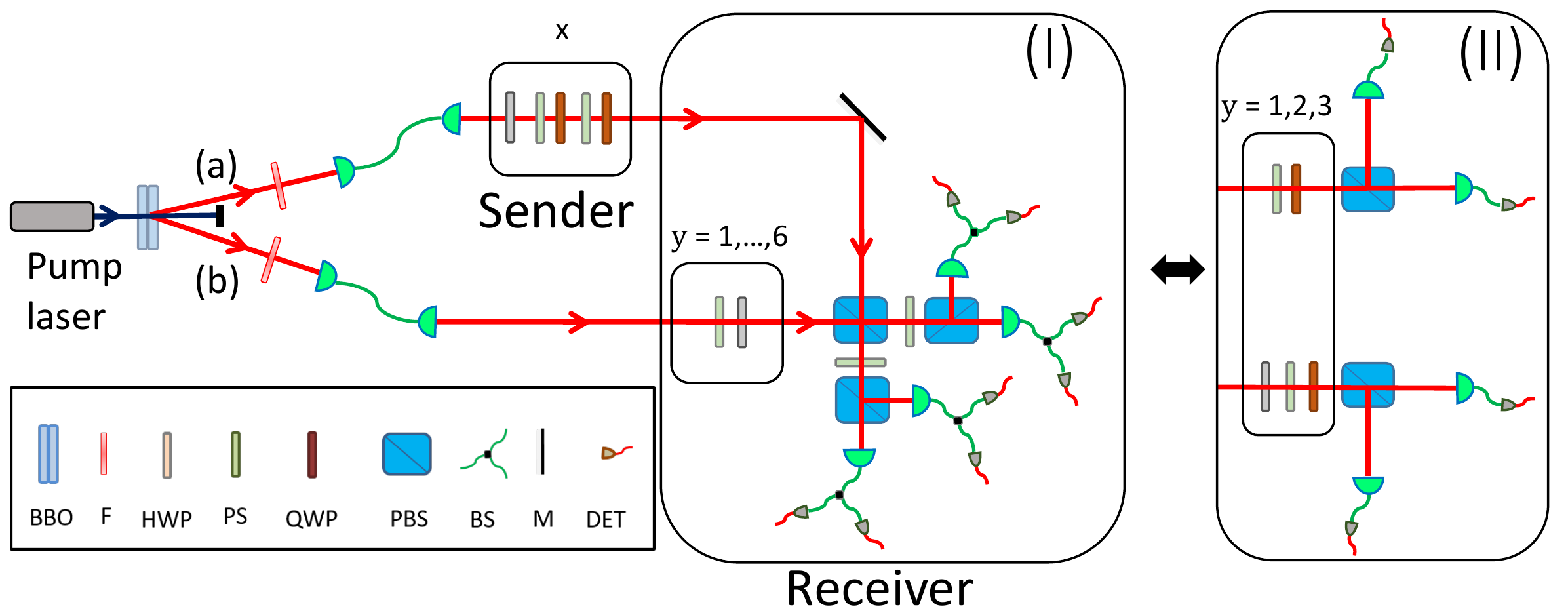}
	\centering
	\caption{\textit{Experimental setup.} Ultraviolet light centred at a wavelength of 390\,nm is focused onto two 2\,mm thick $\beta$ barium borate (BBO) nonlinear crystals placed in cross-configuration to produce photon pairs emitted into two spatial modes $(a)$ and $(b)$ through the second order degenerate type-I SPDC process. The spatial, spectral and  temporal distinguishability between the down-converted photons is carefully removed by  coupling  to single mode  fiber, narrow Filter (F) and quartz wedges respectively and prepare $|\phi^+\rangle$. The unitaries of the sender and receiver are  implemented  using combination of half wave plates (HWP), quarter wave plates (QWP) and phase shifters (PS). (I) The partial Bell state measurements are implemented through two-photon interference, using PBS and HWP plates set at $22.5^\circ$. Beam splitters (BS) are introduced to estimate the projection probabilities before single photon detectors (actively quenched Si-avalanche photodiodes, DET). Outcome $ b = +1$  corresponds to projection onto  $|\phi^+\rangle$, and  outcome $ b = -1$    corresponds to the other Bell states  $|\psi^-\rangle$,   $|\psi^+\rangle$,   and $|\phi^-\rangle$. In (II) partial Bell state measurement is replaced by product polarisation measurements and are performed by using HWPs, QWP and PBSs. Outcome $b \equiv b_1b_2$, with $b_1$,$b_2 \in \{+1,-1\}^2$ corresponds to HH/VV or HV/VH detection when $b=+1$ or $b=-1$ respectively.\label{Fig2}}
\end{figure*}

\subsection{Beyond two-bits models with a partial Bell state analyser}
Consider a communication task with the minimal number of preparations needed to possibly beat two-bit protocols: the sender holds data $x\in\{1,\ldots,5\}$ and the receiver selects questions $y\in\{1,\ldots,6\}$, each with a binary answer $b\in\{+1,-1\}$. Clearly, each question can only yield  partial knowledge about $x$. We consider a simple figure of merit, $\mathcal{S}$, in which each question either has precisely one correct answer or no correct answer. We can rephrase this in terms of the answer ``$b=+1$'' either being awarded one point (if correct), being penalised by one point (if incorrect) or being ignored. Our figure of merit is
\begin{eqnarray}\label{task}
\mathcal{S}\equiv \sum_{x=1}^{5}\sum_{y=1}^6 c_{xy}p(b=+1|x,y) ~,
\end{eqnarray}
where the points awarded for each question are given by
\begin{equation} \label{coeficients}
c=    \left(\begin{array}{cccccc} 1 & 1 & 1 & 0 & 0 & 0\\ -1 & 0 & 0 & 1 & 0 & 0\\ -1 & 0 & 0 & -1 & 1 & 0\\ 0 & -1 & 0 & -1 & -1 & 1\\ 0 & 0 & -1 & -1 & -1 & -1 \end{array}\right).
\end{equation}
This figure of merit comes with favourable properties, but may be viewed as a proof-of-principle construction.

Using two bits of communication, the optimal score is $\mathcal{S}_\text{2 bits}=5$ (see Methods).  To saturate it, the two bits can be encoded as $\{x=1,x=2 \vee 3,x=4,x=5\}$. Indeed,  dense coding substantially improves on the best standard classical protocol, based on one bit of communication, (at best $\mathcal{S}_\text{bit}=3$). It also improves on the best protocol when one bit of communication is assisted by any amount of shared entanglement, specifically $\mathcal{S}_\text{ent+bit}\approx 3.799$ (see Methods). %This is obtained via a numerical search over quantum models that is then matched with an upper bound obtained from the semidefinite programming hierarchy developed in \cite{Tavakoli2021} which in turn uses the concept of informationally-restricted quantum correlations \cite{Tavakoli2020informationally, Tavakoli2020}.

However we can beat the two-bit limit: this time using only simple entangled measurements  that only discriminate one of the four Bell states. Let the sender perform  the following unitaries on her share of $\ket{\phi^+}$, 
\begin{align}\nonumber
&U^\text{S}_1=\openone, \quad U^\text{S}_2=\frac{-\sigma_Z\sqrt{3}-\sigma_X}{2}, \quad U^\text{S}_3=\frac{\sigma_X\sqrt{3}-\sigma_Z}{2},\\
&\quad \quad \quad U^\text{S}_4=\frac{\openone-i\sigma_Y\sqrt{3}}{2},\quad U^\text{S}_5=\frac{\openone+i\sigma_Y\sqrt{3}}{2}.
\end{align} 
These are rotations in the $XZ$-plane of the Bloch sphere. Once the qubit is relayed, the receiver holds the state $U^\text{S}_x\otimes \openone \ket{\phi^+}$. The receiver performs a binary-outcome measurement $\{\ketbra{E_{y}}{E_{y}},\openone-\ketbra{E_{y}}{E_{y}}\}$, where the outcome $b=+1$ corresponds to a projection onto the state $\ket{E_{y}}=U^\text{R}_y\otimes \openone \ket{\phi^+}$ for some unitary $U^\text{R}_y$. Such a measurement may be viewed as a locally rotated partial Bell state analyser; it attempts to discriminate the maximally entangled state $\ket{E_{y}}$ from its orthogonal complement. We choose the unitaries of the receiver as
\begin{align}\nonumber
&U^\text{R}_1=\openone, \quad U^\text{R}_2=\frac{\nu_+\openone+i\nu_-\sigma_Y}{2\sqrt{2}}, \quad U^\text{R}_3=\frac{\nu_+\openone-i\nu_-\sigma_Y}{2\sqrt{2}},\\
& \qquad \quad U^\text{R}_4= U^\text{S}_2, \quad U^\text{R}_5=U^\text{S}_3,\quad U^\text{R}_6=\frac{\openone-i\sigma_Y}{\sqrt{2}} \, .
\end{align} 
where $\nu_\pm=\sqrt{3}\pm1$. Once again, these are rotations in the $XZ$-plane. The figure of merit becomes
\begin{align}\nonumber\label{value}
\mathcal{S}=& \sum_{x,y} c_{xy}|\bracket{E_y}{U^\text{S}_x\otimes \openone}{\phi^+}|^2\\
&=\frac{1}{2}\Tr\left[\left(U^\text{R}_y\right)^\dagger U^\text{S}_x\right]=3+\frac{3\sqrt{3}}{2}\approx 5.598 \, ,
\end{align}
which considerably exceeds the two-bit limit. We note that the value \eqref{value} can be somewhat further increased. Using a numerical search, we find a protocol achieving $\mathcal{S}\approx 5.641$. However, this protocol is of lesser interest since it requires more complicated measurements.  Our above protocol is also robust to unavoidable implementational imperfections. For instance, if the EPR pair is exposed to isotropic noise, so that the state becomes $v\ketbra{\phi^+}{\phi^+}+\frac{1-v}{4}\openone$, for some visibility $v\in[0,1]$, the advantage over dense coding is maintained whenever  $v>\frac{16}{12+3\sqrt{3}}\approx 93\%$.

\begin{figure}[t!]
  \centering
  \includegraphics[width=0.9\columnwidth]{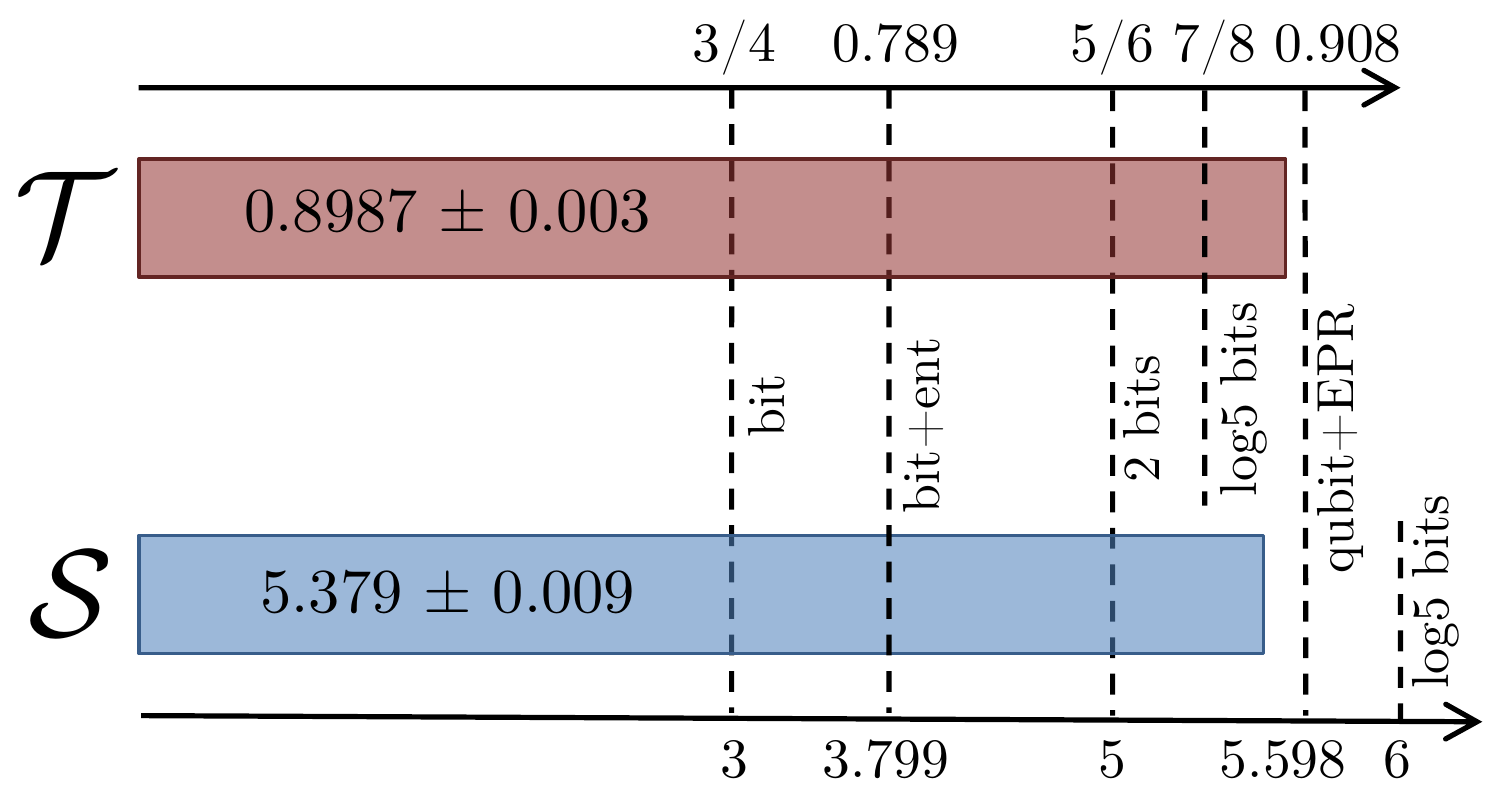}
    \caption{\textit{Experimental results.} Illustration of the experimentally measured performance of the communication tasks and their comparison to the best conceivable protocols based on one bit of classical communication, one bit of classical communication assisted by unbounded entanglement, two bits of classical communication and five-valued classical communication. The two-bit bound is equal to the maximum attainable value using a dense coding protocol. Another dashed line represents the theoretical value of the targeted quantum protocol, based on a shared EPR pair and a communicated qubit.\label{Fig3} }
  \end{figure}

Thanks to the simplicity of the measurements, the correlation advantage can be demonstrated using standard linear optics for polarisation qubits. Using a spontaneous parametric down-conversion (SPDC) process, we  prepare two-photon polarisation entangled state $|\Psi\rangle =  \frac{1}{\sqrt{2}}(|HH\rangle +  |VV\rangle)$. The single-qubit unitaries are implemented using combinations of wave-plates and phase shifters while the partial Bell state measurement is implemented by interfering the two photons via a polarising beam splitter. The setup is illustrated in Fig.~\ref{Fig2} and the specific settings are given in Supplementary Note 2. This leads us to the experimentally measured value of $\mathcal{S}=5.379 \pm 0.009$, which outperforms the dense coding limit by approximately $40$ standard deviations (see Methods). Due to the sizeable violation and the large number of collected events, the $p$-value associated to our falsification of a two-bit model is vanishingly small (see Methods). The result and its relation to the various theoretical limits is illustrated in Fig.~\ref{Fig3}.

However, even though partial Bell state measurements offer sizeable advantages over two-bit protocols in relatively simple photonic experiments, they do not offer a clear path to scalability in terms of dimension or particle number. Also, from a conceptual standpoint, it is not self-evident that entanglement in the measurement must, in general, be indispensable for correlation advantages. To address both the practical and conceptual question, we investigate the possibility of using even the most elementary class of joint measurements, namely product measurements, for entanglement-assisted communication beyond two-bit protocols.

\subsection{Optimal Quantum Random Access Code with product measurements}
	
Unfortunately, the previously discussed RAC seems to offer no advantage over two-bit classical models when the receiver is restricted to product measurements  However, a variation of it, featuring a different number of inputs and settings, does reveal a distinct advantage. Let the sender hold data $x=x_1x_2x_3\in\{0,1\}^3$ and the receiver uniformly select $y\in\{1,2,3\}$ with the aim of recovering $x_y$. The average success rate is 
\begin{equation}
\mathcal{T}=\frac{1}{24}\sum_{x,y}p(b=x_y|x,y).
\end{equation}
A two-bit protocol achieves at best $\mathcal{T}_\text{2 bit}=\frac{5}{6}$ which can be shown considering all vertices of the classical polytope (see Methods). It is saturated by the sender relaying both the majority bit in $x$ and the $x_3$ to the receiver.  Now consider that the sender performs the unitaries
\begin{equation}
U_x=(-1)^{x_1}
\begin{pmatrix}
-\alpha_{x_1}\mu_{x_2x_3} & (-1)^{x_2+x_3}\alpha_{\bar{x}_1}\mu_{x_2x_3}\\
(-1)^{x_2+x_3}\sqrt{2}\alpha_{\bar{x}_1} & \sqrt{2}\alpha_{x_1},
\end{pmatrix}
\end{equation}
where $\mu_{x_2x_3}=(-1)^{x_2}+i(-1)^{x_3}$, $\alpha_s=\frac{1}{2}\sqrt{1+(-1)^s\sqrt{2/3}}$ and the bar-sign denotes bit-flip. The receiver measures three product observables 
\begin{align}
&E_1=\sigma_Z\otimes \sigma_Z \\
&E_2=\frac{1}{2}\sigma_Y\otimes (\sqrt{3}\sigma_Y+\sigma_Z) \\
& E_3=\frac{1}{2} \sigma_X \otimes (\sqrt{3}\sigma_Y-\sigma_Z),
\end{align}
where $E_y\equiv E_{1|y}-E_{2|y}$. This leads to $\mathcal{T}=\frac{1}{2}+\frac{1}{\sqrt{6}}>\mathcal{T}_\text{2 bit}$. In fact, this outperforms  even protocols based on sending five classical symbols. Interestingly, no better quantum protocol is possible, even if based on general entangled measurements. This follows from the fact that the best protocol based on four-dimensional quanta is also known to achieve $\mathcal{T}= \frac{1}{2}+\frac{1}{\sqrt{6}}$ \cite{Navascues2015a}.  Adapting the optical setup (see Figure~\ref{Fig2}), we have demonstrated also this correlation advantage using the same source  and the same measuring time as in the previous experiment. The specific settings are given in Supplementary Note 4. We observe $\mathcal{T}=0.8987\pm 0.003$, which beats the two-bit limit with over 20 standard deviations (see Methods) . The results are illustrated in Figure~\ref{Fig3}.

\section*{Discussion}
The finding, that simple measurements are sufficient for creating quantum correlations that cannot be modelled with two bits of classical communication, and sometimes even constitute an optimal protocol for natural quantum resources, is based on departing from the study of the dense coding task in favour of more general quantum communication tasks. Conceptually, it motivates a research effort into general entanglement-assisted correlations \cite{Tavakoli2021a, Pauwels2022a} based on product measurements for the receiver. A natural question is to determine when and why product measurements are useful for entanglement-assisted quantum communication. It is paired with crucial practical advantages since such protocols circumvents the need for implementing highly demanding entangled measurements in favour of quantum devices that require only single-system measurements. This may make possible both multi-particle and high-dimensional protocols for entanglement-assisted quantum communication that are realistically implementable. It may also offer a viable practical path to otherwise demanding foundational experiments based on these natural quantum resources.

\section*{Methods}

\subsection*{Correlation bounds} When communication is classical and no entanglement is present, $p(b|x,y)$ can be geometrically represented as a polytope whose vertices correspond to deterministic encoding and decoding schemes \cite{Gallego2010}. Consequently, the optimal performance of any linear figure of merit, e.g.~that in Eq.~\eqref{task}, is necessarily attained at a vertex of this polytope. One can thus check the value of the figure of merit at all vertices and select the largest value.  However, when (potentially unbounded) entanglement is added, this picture breaks down. Instead, upper bounds on $\mathcal{S}$ and $\mathcal{T}$ can be determined using the hierarchy of semidefinite programming relaxations developed in \cite{Tavakoli2021a}, which uses the concept of informationally-restricted quantum correlations \cite{InfoCorrelations1, InfoCorrelations2}. Using this method, and matching it with an explicit entanglement-based strategy with classical communication,  we find  $\mathcal{S}_\text{ent+bit}\approx 3.799$ and $\mathcal{T}_\text{ent+bit}\approx 0.789$. 

The optimality of our protocols for the two RACs, corresponding to $\mathcal{R}$ and $\mathcal{T}$, respectively, can be shown as follows. For both results, we use that the set of correlations attainable with an EPR pair and a qubit message is a subset of the set of correlations attainable in scenarios in which the sender and receiver only share classical randomness and communicate a four-dimensional quantum system. In these standard scenarios, it is known that $\mathcal{R}\leq\frac{3}{4}$  \cite{Tavakoli2015} and that $\mathcal{T}\leq \frac{1}{2}+\frac{1}{\sqrt{6}}$ \cite{Navascues} for general protocols. As our protocols saturate these bounds, optimality follows.

\subsection*{Experimental errors} 
To reduce the multi-photons pairs emission we worked at a low rate ($\approx 2500$ two-photon coincidences per sec, ca.~$13\%$ of the singles rate) and increased the measurement time to reduce statistical errors.  We benchmark the state preparation by measuring an average visibility of $0.992\pm 0.001$ in the diagonal polarisation basis. Similarly, we benchmark the two-photon interference by a two-fold Hong-Ou-Mandel dip visibility of $0.961 \pm 0.002$ (See Supplementary Note 7).  For each setting $x$ and $y$, we collect on average 18 million events during a  measurement  time  of  two hours.  The probabilities $p(b|x,y)$ are estimated from the relative frequencies  (see Supplementary Note 3 and 5). The impact of systematic errors was estimated using Monte Carlo simulation. These were reduced by using computerised high precision mounts. (See details in Supplementary Note 6). The experiment is performed using the fair sampling assumption.

\subsection*{Statistical significance}  To express the statistical significance of our experimental results, we follow an approach similar to \cite{Pironio2010} introduced by \cite{Gill2002}, to which we refer for details.  Consider the random variable 
\begin{equation}
\hat{\mathcal{S}}_i = \sum_{xy} c_{xy}\frac{ \chi(b_i=1,x_i=x,y_i=y)}{p(x,y)}, 
\end{equation}
where $i$ corresponds to the $i$th experimental run, $\chi(e)$ is the indicator function for the event $e$, i.e. $\chi(e)= 1$ if the event is observed and $\chi(e) = 0$ otherwise. For our experiment we simply chose $p(x,y) = 1/(6 \times 5) =1/30.$
The random variable $\hat{\mathcal{S}}_i$ may depend on past events, $j<i$, but not on future events, $j>i$. We define $\hat{\mathcal{S}} = \frac{1}{N} \sum_{i=1}^N \hat{\mathcal{S}}_i$ as our estimator for the value of our scoring function $\mathcal{S}$, where $N \sim 18 \times 5 \times 6 $  million is the total number of experimental rounds.

The Azuma-Hoeffding inequality implies that the probability $p$ that dense coding or, equivalently, a two bit communication model will yield a value of $\mathcal{S}$ greater or equal to the observed value is bounded by

\begin{equation}
p\Bigg(\frac{1}{N} \sum_{i=1}^N \hat{\mathcal{S}}_i \geq \mathcal{S}_{\rm 2 bits} + \mu \Bigg)\leq \exp \Bigg(\frac{2N\mu^2}{(c+T)^2} \Bigg),
\end{equation}
where $\mu=0.379$ is the observed violation of the two bit bound, $T=9$ is the classical $2$-bit bound on $-\mathcal{S}$ and $c\equiv \text{max}_{xy}c_{xy}/p(x,y)$. One finds that this probability is vanishingly small. The analogous procedure applies to the data analysis based on the measured value of $\mathcal{T}$.

\section*{Acknowledgements}	
We thank Stefano Pironio and Erik Woodhead for discussions and comments. This work was supported by the Swedish research council, the Knut and Alice Wallenberg Foundation through the Wallenberg Center for Quantum Technology (WACQT), the Wenner-Gren Foundation, the Fonds de la Recherche Scientifique (F.R.S.) – FNRS under Grant No PDR T.0171.22, and under grant Pint-Multi R.8014.21 as part of the QuantERA ERA-NET EU programme,  the FWO and the F.R.S.-FNRS under grant 40007526 of the Excellence of Science (EOS) programme and the F.R.S.-FNRS under a FRIA grant.
\\
\textbf{Data \& code availability:} Source data is provided as supplementary to this paper.  Any additional data and computer codes related to the findings of this paper is available upon request.
\\
\textbf{Author contributions:} A.~T proposed the project. J.~P and A.~T developed the theory. S.~M designed the experiment.  A.~P and E.~H performed the experiments and analysed the data. S.~M and  M.~B supervised the experiment work. All authors contributed to the writing of the manuscript. 
\\
\textbf{Competing interests:} The authors declare no competing interests.

\bibliography{references.bib}

\begin{thebibliography}{39}%
\makeatletter
\providecommand \@ifxundefined [1]{%
 \@ifx{#1\undefined}
}%
\providecommand \@ifnum [1]{%
 \ifnum #1\expandafter \@firstoftwo
 \else \expandafter \@secondoftwo
 \fi
}%
\providecommand \@ifx [1]{%
 \ifx #1\expandafter \@firstoftwo
 \else \expandafter \@secondoftwo
 \fi
}%
\providecommand \natexlab [1]{#1}%
\providecommand \enquote  [1]{``#1''}%
\providecommand \bibnamefont  [1]{#1}%
\providecommand \bibfnamefont [1]{#1}%
\providecommand \citenamefont [1]{#1}%
\providecommand \href@noop [0]{\@secondoftwo}%
\providecommand \href [0]{\begingroup \@sanitize@url \@href}%
\providecommand \@href[1]{\@@startlink{#1}\@@href}%
\providecommand \@@href[1]{\endgroup#1\@@endlink}%
\providecommand \@sanitize@url [0]{\catcode `\\12\catcode `\$12\catcode
  `\&12\catcode `\#12\catcode `\^12\catcode `\_12\catcode `\%12\relax}%
\providecommand \@@startlink[1]{}%
\providecommand \@@endlink[0]{}%
\providecommand \url  [0]{\begingroup\@sanitize@url \@url }%
\providecommand \@url [1]{\endgroup\@href {#1}{\urlprefix }}%
\providecommand \urlprefix  [0]{URL }%
\providecommand \Eprint [0]{\href }%
\providecommand \doibase [0]{https://doi.org/}%
\providecommand \selectlanguage [0]{\@gobble}%
\providecommand \bibinfo  [0]{\@secondoftwo}%
\providecommand \bibfield  [0]{\@secondoftwo}%
\providecommand \translation [1]{[#1]}%
\providecommand \BibitemOpen [0]{}%
\providecommand \bibitemStop [0]{}%
\providecommand \bibitemNoStop [0]{.\EOS\space}%
\providecommand \EOS [0]{\spacefactor3000\relax}%
\providecommand \BibitemShut  [1]{\csname bibitem#1\endcsname}%
\let\auto@bib@innerbib\@empty
%</preamble>
\bibitem [{\citenamefont {Hensen}\ \emph {et~al.}(2015)\citenamefont {Hensen},
  \citenamefont {Bernien}, \citenamefont {Dr{\'{e}}au}, \citenamefont
  {Reiserer}, \citenamefont {Kalb}, \citenamefont {Blok}, \citenamefont
  {Ruitenberg}, \citenamefont {Vermeulen}, \citenamefont {Schouten},
  \citenamefont {Abell{\'{a}}n}, \citenamefont {Amaya}, \citenamefont
  {Pruneri}, \citenamefont {Mitchell}, \citenamefont {Markham}, \citenamefont
  {Twitchen}, \citenamefont {Elkouss}, \citenamefont {Wehner}, \citenamefont
  {Taminiau},\ and\ \citenamefont {Hanson}}]{Hensen2015}%
  \BibitemOpen
  \bibfield  {author} {\bibinfo {author} {\bibfnamefont {B.}~\bibnamefont
  {Hensen}}, \bibinfo {author} {\bibfnamefont {H.}~\bibnamefont {Bernien}},
  \bibinfo {author} {\bibfnamefont {A.~E.}\ \bibnamefont {Dr{\'{e}}au}},
  \bibinfo {author} {\bibfnamefont {A.}~\bibnamefont {Reiserer}}, \bibinfo
  {author} {\bibfnamefont {N.}~\bibnamefont {Kalb}}, \bibinfo {author}
  {\bibfnamefont {M.~S.}\ \bibnamefont {Blok}}, \bibinfo {author}
  {\bibfnamefont {J.}~\bibnamefont {Ruitenberg}}, \bibinfo {author}
  {\bibfnamefont {R.~F.~L.}\ \bibnamefont {Vermeulen}}, \bibinfo {author}
  {\bibfnamefont {R.~N.}\ \bibnamefont {Schouten}}, \bibinfo {author}
  {\bibfnamefont {C.}~\bibnamefont {Abell{\'{a}}n}}, \bibinfo {author}
  {\bibfnamefont {W.}~\bibnamefont {Amaya}}, \bibinfo {author} {\bibfnamefont
  {V.}~\bibnamefont {Pruneri}}, \bibinfo {author} {\bibfnamefont {M.~W.}\
  \bibnamefont {Mitchell}}, \bibinfo {author} {\bibfnamefont {M.}~\bibnamefont
  {Markham}}, \bibinfo {author} {\bibfnamefont {D.~J.}\ \bibnamefont
  {Twitchen}}, \bibinfo {author} {\bibfnamefont {D.}~\bibnamefont {Elkouss}},
  \bibinfo {author} {\bibfnamefont {S.}~\bibnamefont {Wehner}}, \bibinfo
  {author} {\bibfnamefont {T.~H.}\ \bibnamefont {Taminiau}},\ and\ \bibinfo
  {author} {\bibfnamefont {R.}~\bibnamefont {Hanson}},\ }\bibfield  {title}
  {\bibinfo {title} {Loophole-free bell inequality violation using electron
  spins separated by 1.3 kilometres},\ }\href
  {https://doi.org/10.1038/nature15759} {\bibfield  {journal} {\bibinfo
  {journal} {Nature}\ }\textbf {\bibinfo {volume} {526}},\ \bibinfo {pages}
  {682} (\bibinfo {year} {2015})}\BibitemShut {NoStop}%
\bibitem [{\citenamefont {Giustina}\ \emph {et~al.}(2015)\citenamefont
  {Giustina}, \citenamefont {Versteegh}, \citenamefont {Wengerowsky},
  \citenamefont {Handsteiner}, \citenamefont {Hochrainer}, \citenamefont
  {Phelan}, \citenamefont {Steinlechner}, \citenamefont {Kofler}, \citenamefont
  {Larsson}, \citenamefont {Abell\'an}, \citenamefont {Amaya}, \citenamefont
  {Pruneri}, \citenamefont {Mitchell}, \citenamefont {Beyer}, \citenamefont
  {Gerrits}, \citenamefont {Lita}, \citenamefont {Shalm}, \citenamefont {Nam},
  \citenamefont {Scheidl}, \citenamefont {Ursin}, \citenamefont {Wittmann},\
  and\ \citenamefont {Zeilinger}}]{Zeilinger2015}%
  \BibitemOpen
  \bibfield  {author} {\bibinfo {author} {\bibfnamefont {M.}~\bibnamefont
  {Giustina}}, \bibinfo {author} {\bibfnamefont {M.~A.~M.}\ \bibnamefont
  {Versteegh}}, \bibinfo {author} {\bibfnamefont {S.}~\bibnamefont
  {Wengerowsky}}, \bibinfo {author} {\bibfnamefont {J.}~\bibnamefont
  {Handsteiner}}, \bibinfo {author} {\bibfnamefont {A.}~\bibnamefont
  {Hochrainer}}, \bibinfo {author} {\bibfnamefont {K.}~\bibnamefont {Phelan}},
  \bibinfo {author} {\bibfnamefont {F.}~\bibnamefont {Steinlechner}}, \bibinfo
  {author} {\bibfnamefont {J.}~\bibnamefont {Kofler}}, \bibinfo {author}
  {\bibfnamefont {J.-A.}\ \bibnamefont {Larsson}}, \bibinfo {author}
  {\bibfnamefont {C.}~\bibnamefont {Abell\'an}}, \bibinfo {author}
  {\bibfnamefont {W.}~\bibnamefont {Amaya}}, \bibinfo {author} {\bibfnamefont
  {V.}~\bibnamefont {Pruneri}}, \bibinfo {author} {\bibfnamefont {M.~W.}\
  \bibnamefont {Mitchell}}, \bibinfo {author} {\bibfnamefont {J.}~\bibnamefont
  {Beyer}}, \bibinfo {author} {\bibfnamefont {T.}~\bibnamefont {Gerrits}},
  \bibinfo {author} {\bibfnamefont {A.~E.}\ \bibnamefont {Lita}}, \bibinfo
  {author} {\bibfnamefont {L.~K.}\ \bibnamefont {Shalm}}, \bibinfo {author}
  {\bibfnamefont {S.~W.}\ \bibnamefont {Nam}}, \bibinfo {author} {\bibfnamefont
  {T.}~\bibnamefont {Scheidl}}, \bibinfo {author} {\bibfnamefont
  {R.}~\bibnamefont {Ursin}}, \bibinfo {author} {\bibfnamefont
  {B.}~\bibnamefont {Wittmann}},\ and\ \bibinfo {author} {\bibfnamefont
  {A.}~\bibnamefont {Zeilinger}},\ }\bibfield  {title} {\bibinfo {title}
  {Significant-loophole-free test of bell's theorem with entangled photons},\
  }\href {https://doi.org/10.1103/PhysRevLett.115.250401} {\bibfield  {journal}
  {\bibinfo  {journal} {Phys. Rev. Lett.}\ }\textbf {\bibinfo {volume} {115}},\
  \bibinfo {pages} {250401} (\bibinfo {year} {2015})}\BibitemShut {NoStop}%
\bibitem [{\citenamefont {Shalm}\ \emph {et~al.}(2015)\citenamefont {Shalm}
  \emph {et~al.}}]{Shalm2015}%
  \BibitemOpen
  \bibfield  {author} {\bibinfo {author} {\bibfnamefont {L.}~\bibnamefont
  {Shalm}} \emph {et~al.},\ }\bibfield  {title} {\bibinfo {title} {Strong
  {{Loophole}}-{{Free Test}} of {{Local Realism}}},\ }\href
  {https://doi.org/10.1103/PhysRevLett.115.250402} {\bibfield  {journal}
  {\bibinfo  {journal} {\prl}\ }\textbf {\bibinfo {volume} {115}},\ \bibinfo
  {pages} {250402} (\bibinfo {year} {2015})}\BibitemShut {NoStop}%
\bibitem [{\citenamefont {Rosenfeld}\ \emph {et~al.}(2017)\citenamefont
  {Rosenfeld}, \citenamefont {Burchardt}, \citenamefont {Garthoff},
  \citenamefont {Redeker}, \citenamefont {Ortegel}, \citenamefont {Rau},\ and\
  \citenamefont {Weinfurter}}]{Rosenfeld2017}%
  \BibitemOpen
  \bibfield  {author} {\bibinfo {author} {\bibfnamefont {W.}~\bibnamefont
  {Rosenfeld}}, \bibinfo {author} {\bibfnamefont {D.}~\bibnamefont
  {Burchardt}}, \bibinfo {author} {\bibfnamefont {R.}~\bibnamefont {Garthoff}},
  \bibinfo {author} {\bibfnamefont {K.}~\bibnamefont {Redeker}}, \bibinfo
  {author} {\bibfnamefont {N.}~\bibnamefont {Ortegel}}, \bibinfo {author}
  {\bibfnamefont {M.}~\bibnamefont {Rau}},\ and\ \bibinfo {author}
  {\bibfnamefont {H.}~\bibnamefont {Weinfurter}},\ }\bibfield  {title}
  {\bibinfo {title} {Event-ready bell test using entangled atoms simultaneously
  closing detection and locality loopholes},\ }\href
  {https://doi.org/10.1103/PhysRevLett.119.010402} {\bibfield  {journal}
  {\bibinfo  {journal} {\prl}\ }\textbf {\bibinfo {volume} {119}},\ \bibinfo
  {pages} {010402} (\bibinfo {year} {2017})}\BibitemShut {NoStop}%
\bibitem [{\citenamefont {Gallego}\ \emph {et~al.}(2010)\citenamefont
  {Gallego}, \citenamefont {Brunner}, \citenamefont {Hadley},\ and\
  \citenamefont {Ac\'{\i}n}}]{Gallego2010}%
  \BibitemOpen
  \bibfield  {author} {\bibinfo {author} {\bibfnamefont {R.}~\bibnamefont
  {Gallego}}, \bibinfo {author} {\bibfnamefont {N.}~\bibnamefont {Brunner}},
  \bibinfo {author} {\bibfnamefont {C.}~\bibnamefont {Hadley}},\ and\ \bibinfo
  {author} {\bibfnamefont {A.}~\bibnamefont {Ac\'{\i}n}},\ }\bibfield  {title}
  {\bibinfo {title} {Device-independent tests of classical and quantum
  dimensions},\ }\href {https://doi.org/10.1103/PhysRevLett.105.230501}
  {\bibfield  {journal} {\bibinfo  {journal} {Phys. Rev. Lett.}\ }\textbf
  {\bibinfo {volume} {105}},\ \bibinfo {pages} {230501} (\bibinfo {year}
  {2010})}\BibitemShut {NoStop}%
\bibitem [{\citenamefont {Ahrens}\ \emph {et~al.}(2012)\citenamefont {Ahrens},
  \citenamefont {Badziag}, \citenamefont {Cabello},\ and\ \citenamefont
  {Bourennane}}]{Ahrens2012}%
  \BibitemOpen
  \bibfield  {author} {\bibinfo {author} {\bibfnamefont {J.}~\bibnamefont
  {Ahrens}}, \bibinfo {author} {\bibfnamefont {P.}~\bibnamefont {Badziag}},
  \bibinfo {author} {\bibfnamefont {A.}~\bibnamefont {Cabello}},\ and\ \bibinfo
  {author} {\bibfnamefont {M.}~\bibnamefont {Bourennane}},\ }\bibfield  {title}
  {\bibinfo {title} {Experimental device-independent tests of classical and
  quantum dimensions},\ }\href {https://doi.org/10.1038/nphys2333} {\bibfield
  {journal} {\bibinfo  {journal} {Nature Phys.}\ }\textbf {\bibinfo {volume}
  {8}},\ \bibinfo {pages} {592–595} (\bibinfo {year} {2012})}\BibitemShut
  {NoStop}%
\bibitem [{\citenamefont {Hendrych}\ \emph {et~al.}(2012)\citenamefont
  {Hendrych}, \citenamefont {Gallego}, \citenamefont {Mi\v{c}uda},
  \citenamefont {Brunner}, \citenamefont {Ac\'in},\ and\ \citenamefont
  {Torres}}]{Hendrych2012}%
  \BibitemOpen
  \bibfield  {author} {\bibinfo {author} {\bibfnamefont {M.}~\bibnamefont
  {Hendrych}}, \bibinfo {author} {\bibfnamefont {R.}~\bibnamefont {Gallego}},
  \bibinfo {author} {\bibfnamefont {M.}~\bibnamefont {Mi\v{c}uda}}, \bibinfo
  {author} {\bibfnamefont {N.}~\bibnamefont {Brunner}}, \bibinfo {author}
  {\bibfnamefont {A.}~\bibnamefont {Ac\'in}},\ and\ \bibinfo {author}
  {\bibfnamefont {J.~P.}\ \bibnamefont {Torres}},\ }\bibfield  {title}
  {\bibinfo {title} {Experimental estimation of the dimension of classical and
  quantum systems},\ }\href {https://doi.org/10.1038/nphys2334} {\bibfield
  {journal} {\bibinfo  {journal} {Nature Phys.}\ }\textbf {\bibinfo {volume}
  {8}},\ \bibinfo {pages} {588–591} (\bibinfo {year} {2012})}\BibitemShut
  {NoStop}%
\bibitem [{\citenamefont {Bennett}\ and\ \citenamefont
  {Wiesner}(1992)}]{Bennett1992}%
  \BibitemOpen
  \bibfield  {author} {\bibinfo {author} {\bibfnamefont {C.~H.}\ \bibnamefont
  {Bennett}}\ and\ \bibinfo {author} {\bibfnamefont {S.~J.}\ \bibnamefont
  {Wiesner}},\ }\bibfield  {title} {\bibinfo {title} {Communication via one-
  and two-particle operators on {E}instein-{P}odolsky-{R}osen states},\ }\href
  {https://doi.org/10.1103/PhysRevLett.69.2881} {\bibfield  {journal} {\bibinfo
   {journal} {\prl}\ }\textbf {\bibinfo {volume} {69}},\ \bibinfo {pages}
  {2881} (\bibinfo {year} {1992})}\BibitemShut {NoStop}%
\bibitem [{\citenamefont {Holevo}(1973)}]{Holevo1973}%
  \BibitemOpen
  \bibfield  {author} {\bibinfo {author} {\bibfnamefont {A.~S.}\ \bibnamefont
  {Holevo}},\ }\bibfield  {title} {\bibinfo {title} {Bounds for the quantity of
  information transmitted by a quantum communication channel},\ }\href
  {http://mi.mathnet.ru/ppi903} {\bibfield  {journal} {\bibinfo  {journal}
  {Problems Inform. Transmission}\ }\textbf {\bibinfo {volume} {9}},\ \bibinfo
  {pages} {3} (\bibinfo {year} {1973})}\BibitemShut {NoStop}%
\bibitem [{\citenamefont {Bennett}\ \emph {et~al.}(1999)\citenamefont
  {Bennett}, \citenamefont {Shor}, \citenamefont {Smolin},\ and\ \citenamefont
  {Thapliyal}}]{Bennett1999}%
  \BibitemOpen
  \bibfield  {author} {\bibinfo {author} {\bibfnamefont {C.~H.}\ \bibnamefont
  {Bennett}}, \bibinfo {author} {\bibfnamefont {P.~W.}\ \bibnamefont {Shor}},
  \bibinfo {author} {\bibfnamefont {J.~A.}\ \bibnamefont {Smolin}},\ and\
  \bibinfo {author} {\bibfnamefont {A.~V.}\ \bibnamefont {Thapliyal}},\
  }\bibfield  {title} {\bibinfo {title} {Entanglement-assisted classical
  capacity of noisy quantum channels},\ }\href
  {https://doi.org/10.1103/PhysRevLett.83.3081} {\bibfield  {journal} {\bibinfo
   {journal} {\prl}\ }\textbf {\bibinfo {volume} {83}},\ \bibinfo {pages}
  {3081} (\bibinfo {year} {1999})}\BibitemShut {NoStop}%
\bibitem [{\citenamefont {Pauwels}\ \emph
  {et~al.}(2022{\natexlab{a}})\citenamefont {Pauwels}, \citenamefont {Pironio},
  \citenamefont {Cruzeiro},\ and\ \citenamefont {Tavakoli}}]{Pauwels2022}%
  \BibitemOpen
  \bibfield  {author} {\bibinfo {author} {\bibfnamefont {J.}~\bibnamefont
  {Pauwels}}, \bibinfo {author} {\bibfnamefont {S.}~\bibnamefont {Pironio}},
  \bibinfo {author} {\bibfnamefont {E.~Z.}\ \bibnamefont {Cruzeiro}},\ and\
  \bibinfo {author} {\bibfnamefont {A.}~\bibnamefont {Tavakoli}},\ }\bibfield
  {title} {\bibinfo {title} {Adaptive advantage in entanglement-assisted
  communications},\ }\href {https://doi.org/10.1103/PhysRevLett.129.120504}
  {\bibfield  {journal} {\bibinfo  {journal} {Phys. Rev. Lett.}\ }\textbf
  {\bibinfo {volume} {129}},\ \bibinfo {pages} {120504} (\bibinfo {year}
  {2022}{\natexlab{a}})}\BibitemShut {NoStop}%
\bibitem [{\citenamefont {Schaetz}\ \emph {et~al.}(2004)\citenamefont
  {Schaetz}, \citenamefont {Barrett}, \citenamefont {Leibfried}, \citenamefont
  {Chiaverini}, \citenamefont {Britton}, \citenamefont {Itano}, \citenamefont
  {Jost}, \citenamefont {Langer},\ and\ \citenamefont
  {Wineland}}]{Schaetz2004}%
  \BibitemOpen
  \bibfield  {author} {\bibinfo {author} {\bibfnamefont {T.}~\bibnamefont
  {Schaetz}}, \bibinfo {author} {\bibfnamefont {M.~D.}\ \bibnamefont
  {Barrett}}, \bibinfo {author} {\bibfnamefont {D.}~\bibnamefont {Leibfried}},
  \bibinfo {author} {\bibfnamefont {J.}~\bibnamefont {Chiaverini}}, \bibinfo
  {author} {\bibfnamefont {J.}~\bibnamefont {Britton}}, \bibinfo {author}
  {\bibfnamefont {W.~M.}\ \bibnamefont {Itano}}, \bibinfo {author}
  {\bibfnamefont {J.~D.}\ \bibnamefont {Jost}}, \bibinfo {author}
  {\bibfnamefont {C.}~\bibnamefont {Langer}},\ and\ \bibinfo {author}
  {\bibfnamefont {D.~J.}\ \bibnamefont {Wineland}},\ }\bibfield  {title}
  {\bibinfo {title} {Quantum dense coding with atomic qubits},\ }\href
  {https://doi.org/10.1103/PhysRevLett.93.040505} {\bibfield  {journal}
  {\bibinfo  {journal} {Phys. Rev. Lett.}\ }\textbf {\bibinfo {volume} {93}},\
  \bibinfo {pages} {040505} (\bibinfo {year} {2004})}\BibitemShut {NoStop}%
\bibitem [{\citenamefont {Fang}\ \emph {et~al.}(2000)\citenamefont {Fang},
  \citenamefont {Zhu}, \citenamefont {Feng}, \citenamefont {Mao},\ and\
  \citenamefont {Du}}]{Fang2000}%
  \BibitemOpen
  \bibfield  {author} {\bibinfo {author} {\bibfnamefont {X.}~\bibnamefont
  {Fang}}, \bibinfo {author} {\bibfnamefont {X.}~\bibnamefont {Zhu}}, \bibinfo
  {author} {\bibfnamefont {M.}~\bibnamefont {Feng}}, \bibinfo {author}
  {\bibfnamefont {X.}~\bibnamefont {Mao}},\ and\ \bibinfo {author}
  {\bibfnamefont {F.}~\bibnamefont {Du}},\ }\bibfield  {title} {\bibinfo
  {title} {Experimental implementation of dense coding using nuclear magnetic
  resonance},\ }\href {https://doi.org/10.1103/PhysRevA.61.022307} {\bibfield
  {journal} {\bibinfo  {journal} {Phys. Rev. A}\ }\textbf {\bibinfo {volume}
  {61}},\ \bibinfo {pages} {022307} (\bibinfo {year} {2000})}\BibitemShut
  {NoStop}%
\bibitem [{\citenamefont {L\"utkenhaus}\ \emph {et~al.}(1999)\citenamefont
  {L\"utkenhaus}, \citenamefont {Calsamiglia},\ and\ \citenamefont
  {Suominen}}]{Luetkenhaus1999}%
  \BibitemOpen
  \bibfield  {author} {\bibinfo {author} {\bibfnamefont {N.}~\bibnamefont
  {L\"utkenhaus}}, \bibinfo {author} {\bibfnamefont {J.}~\bibnamefont
  {Calsamiglia}},\ and\ \bibinfo {author} {\bibfnamefont {K.-A.}\ \bibnamefont
  {Suominen}},\ }\bibfield  {title} {\bibinfo {title} {{Bell} measurements for
  teleportation},\ }\href {https://doi.org/10.1103/PhysRevA.59.3295} {\bibfield
   {journal} {\bibinfo  {journal} {\pra}\ }\textbf {\bibinfo {volume} {59}},\
  \bibinfo {pages} {3295} (\bibinfo {year} {1999})}\BibitemShut {NoStop}%
\bibitem [{\citenamefont {Vaidman}\ and\ \citenamefont
  {Yoran}(1999)}]{Vaidman1999}%
  \BibitemOpen
  \bibfield  {author} {\bibinfo {author} {\bibfnamefont {L.}~\bibnamefont
  {Vaidman}}\ and\ \bibinfo {author} {\bibfnamefont {N.}~\bibnamefont
  {Yoran}},\ }\bibfield  {title} {\bibinfo {title} {Methods for reliable
  teleportation},\ }\href {https://doi.org/10.1103/PhysRevA.59.116} {\bibfield
  {journal} {\bibinfo  {journal} {Phys. Rev. A}\ }\textbf {\bibinfo {volume}
  {59}},\ \bibinfo {pages} {116} (\bibinfo {year} {1999})}\BibitemShut
  {NoStop}%
\bibitem [{\citenamefont {Kwiat}\ and\ \citenamefont
  {Weinfurter}(1998)}]{Kwiat1998}%
  \BibitemOpen
  \bibfield  {author} {\bibinfo {author} {\bibfnamefont {P.~G.}\ \bibnamefont
  {Kwiat}}\ and\ \bibinfo {author} {\bibfnamefont {H.}~\bibnamefont
  {Weinfurter}},\ }\bibfield  {title} {\bibinfo {title} {Embedded bell-state
  analysis},\ }\href {https://doi.org/10.1103/PhysRevA.58.R2623} {\bibfield
  {journal} {\bibinfo  {journal} {Phys. Rev. A}\ }\textbf {\bibinfo {volume}
  {58}},\ \bibinfo {pages} {R2623} (\bibinfo {year} {1998})}\BibitemShut
  {NoStop}%
\bibitem [{\citenamefont {Mattle}\ \emph {et~al.}(1996)\citenamefont {Mattle},
  \citenamefont {Weinfurter}, \citenamefont {Kwiat},\ and\ \citenamefont
  {Zeilinger}}]{Mattle1996}%
  \BibitemOpen
  \bibfield  {author} {\bibinfo {author} {\bibfnamefont {K.}~\bibnamefont
  {Mattle}}, \bibinfo {author} {\bibfnamefont {H.}~\bibnamefont {Weinfurter}},
  \bibinfo {author} {\bibfnamefont {P.~G.}\ \bibnamefont {Kwiat}},\ and\
  \bibinfo {author} {\bibfnamefont {A.}~\bibnamefont {Zeilinger}},\ }\bibfield
  {title} {\bibinfo {title} {Dense coding in experimental quantum
  communication},\ }\href {https://doi.org/10.1103/PhysRevLett.76.4656}
  {\bibfield  {journal} {\bibinfo  {journal} {\prl}\ }\textbf {\bibinfo
  {volume} {76}},\ \bibinfo {pages} {4656} (\bibinfo {year}
  {1996})}\BibitemShut {NoStop}%
\bibitem [{\citenamefont {Braunstein}\ and\ \citenamefont
  {Kimble}(2000)}]{Braunstein2000}%
  \BibitemOpen
  \bibfield  {author} {\bibinfo {author} {\bibfnamefont {S.~L.}\ \bibnamefont
  {Braunstein}}\ and\ \bibinfo {author} {\bibfnamefont {H.~J.}\ \bibnamefont
  {Kimble}},\ }\bibfield  {title} {\bibinfo {title} {Dense coding for
  continuous variables},\ }\href {https://doi.org/10.1103/PhysRevA.61.042302}
  {\bibfield  {journal} {\bibinfo  {journal} {Phys. Rev. A}\ }\textbf {\bibinfo
  {volume} {61}},\ \bibinfo {pages} {042302} (\bibinfo {year}
  {2000})}\BibitemShut {NoStop}%
\bibitem [{\citenamefont {Li}\ \emph {et~al.}(2002)\citenamefont {Li},
  \citenamefont {Pan}, \citenamefont {Jing}, \citenamefont {Zhang},
  \citenamefont {Xie},\ and\ \citenamefont {Peng}}]{Li2002}%
  \BibitemOpen
  \bibfield  {author} {\bibinfo {author} {\bibfnamefont {X.}~\bibnamefont
  {Li}}, \bibinfo {author} {\bibfnamefont {Q.}~\bibnamefont {Pan}}, \bibinfo
  {author} {\bibfnamefont {J.}~\bibnamefont {Jing}}, \bibinfo {author}
  {\bibfnamefont {J.}~\bibnamefont {Zhang}}, \bibinfo {author} {\bibfnamefont
  {C.}~\bibnamefont {Xie}},\ and\ \bibinfo {author} {\bibfnamefont
  {K.}~\bibnamefont {Peng}},\ }\bibfield  {title} {\bibinfo {title} {Quantum
  dense coding exploiting a bright einstein-podolsky-rosen beam},\ }\href
  {https://doi.org/10.1103/PhysRevLett.88.047904} {\bibfield  {journal}
  {\bibinfo  {journal} {\prl}\ }\textbf {\bibinfo {volume} {88}},\ \bibinfo
  {pages} {047904} (\bibinfo {year} {2002})}\BibitemShut {NoStop}%
\bibitem [{\citenamefont {Schuck}\ \emph {et~al.}(2006)\citenamefont {Schuck},
  \citenamefont {Huber}, \citenamefont {Kurtsiefer},\ and\ \citenamefont
  {Weinfurter}}]{Schuck2006}%
  \BibitemOpen
  \bibfield  {author} {\bibinfo {author} {\bibfnamefont {C.}~\bibnamefont
  {Schuck}}, \bibinfo {author} {\bibfnamefont {G.}~\bibnamefont {Huber}},
  \bibinfo {author} {\bibfnamefont {C.}~\bibnamefont {Kurtsiefer}},\ and\
  \bibinfo {author} {\bibfnamefont {H.}~\bibnamefont {Weinfurter}},\ }\bibfield
   {title} {\bibinfo {title} {Complete deterministic linear optics bell state
  analysis},\ }\href {https://doi.org/10.1103/PhysRevLett.96.190501} {\bibfield
   {journal} {\bibinfo  {journal} {\prl}\ }\textbf {\bibinfo {volume} {96}},\
  \bibinfo {pages} {190501} (\bibinfo {year} {2006})}\BibitemShut {NoStop}%
\bibitem [{\citenamefont {Barbieri}\ \emph {et~al.}(2007)\citenamefont
  {Barbieri}, \citenamefont {Vallone}, \citenamefont {Mataloni},\ and\
  \citenamefont {De~Martini}}]{Barbieri2007}%
  \BibitemOpen
  \bibfield  {author} {\bibinfo {author} {\bibfnamefont {M.}~\bibnamefont
  {Barbieri}}, \bibinfo {author} {\bibfnamefont {G.}~\bibnamefont {Vallone}},
  \bibinfo {author} {\bibfnamefont {P.}~\bibnamefont {Mataloni}},\ and\
  \bibinfo {author} {\bibfnamefont {F.}~\bibnamefont {De~Martini}},\ }\bibfield
   {title} {\bibinfo {title} {Complete and deterministic discrimination of
  polarization bell states assisted by momentum entanglement},\ }\href
  {https://doi.org/10.1103/PhysRevA.75.042317} {\bibfield  {journal} {\bibinfo
  {journal} {Phys. Rev. A}\ }\textbf {\bibinfo {volume} {75}},\ \bibinfo
  {pages} {042317} (\bibinfo {year} {2007})}\BibitemShut {NoStop}%
\bibitem [{\citenamefont {Barreiro}\ \emph {et~al.}(2008)\citenamefont
  {Barreiro}, \citenamefont {Wei},\ and\ \citenamefont {Kwiat}}]{Barreiro2008}%
  \BibitemOpen
  \bibfield  {author} {\bibinfo {author} {\bibfnamefont {J.~T.}\ \bibnamefont
  {Barreiro}}, \bibinfo {author} {\bibfnamefont {T.-C.}\ \bibnamefont {Wei}},\
  and\ \bibinfo {author} {\bibfnamefont {P.~G.}\ \bibnamefont {Kwiat}},\
  }\bibfield  {title} {\bibinfo {title} {Beating the channel capacity limit for
  linear photonic superdense coding},\ }\href
  {https://doi.org/10.1038/nphys919} {\bibfield  {journal} {\bibinfo  {journal}
  {Nature Phys.}\ }\textbf {\bibinfo {volume} {4}},\ \bibinfo {pages} {282}
  (\bibinfo {year} {2008})}\BibitemShut {NoStop}%
\bibitem [{\citenamefont {Williams}\ \emph {et~al.}(2017)\citenamefont
  {Williams}, \citenamefont {Sadlier},\ and\ \citenamefont
  {Humble}}]{Williams2017}%
  \BibitemOpen
  \bibfield  {author} {\bibinfo {author} {\bibfnamefont {B.~P.}\ \bibnamefont
  {Williams}}, \bibinfo {author} {\bibfnamefont {R.~J.}\ \bibnamefont
  {Sadlier}},\ and\ \bibinfo {author} {\bibfnamefont {T.~S.}\ \bibnamefont
  {Humble}},\ }\bibfield  {title} {\bibinfo {title} {Superdense coding over
  optical fiber links with complete bell-state measurements},\ }\href
  {https://doi.org/10.1103/PhysRevLett.118.050501} {\bibfield  {journal}
  {\bibinfo  {journal} {\prl}\ }\textbf {\bibinfo {volume} {118}},\ \bibinfo
  {pages} {050501} (\bibinfo {year} {2017})}\BibitemShut {NoStop}%
\bibitem [{\citenamefont {Hu}\ \emph {et~al.}(2018)\citenamefont {Hu},
  \citenamefont {Guo}, \citenamefont {Liu}, \citenamefont {Huang},
  \citenamefont {Li},\ and\ \citenamefont {Guo}}]{Hu2018}%
  \BibitemOpen
  \bibfield  {author} {\bibinfo {author} {\bibfnamefont {X.-M.}\ \bibnamefont
  {Hu}}, \bibinfo {author} {\bibfnamefont {Y.}~\bibnamefont {Guo}}, \bibinfo
  {author} {\bibfnamefont {B.-H.}\ \bibnamefont {Liu}}, \bibinfo {author}
  {\bibfnamefont {Y.-F.}\ \bibnamefont {Huang}}, \bibinfo {author}
  {\bibfnamefont {C.-F.}\ \bibnamefont {Li}},\ and\ \bibinfo {author}
  {\bibfnamefont {G.-C.}\ \bibnamefont {Guo}},\ }\bibfield  {title} {\bibinfo
  {title} {Beating the channel capacity limit for superdense coding with
  entangled ququarts},\ }\href {https://doi.org/10.1126/sciadv.aat9304}
  {\bibfield  {journal} {\bibinfo  {journal} {Science Advances}\ }\textbf
  {\bibinfo {volume} {4}},\ \bibinfo {pages} {eaat9304} (\bibinfo {year}
  {2018})}\BibitemShut {NoStop}%
\bibitem [{\citenamefont {Wiesner}(1983)}]{Wiesner1983}%
  \BibitemOpen
  \bibfield  {author} {\bibinfo {author} {\bibfnamefont {S.}~\bibnamefont
  {Wiesner}},\ }\bibfield  {title} {\bibinfo {title} {Conjugate coding},\
  }\href {https://doi.org/10.1145/1008908.1008920} {\bibfield  {journal}
  {\bibinfo  {journal} {SIGACT News}\ }\textbf {\bibinfo {volume} {15}},\
  \bibinfo {pages} {78–88} (\bibinfo {year} {1983})}\BibitemShut {NoStop}%
\bibitem [{\citenamefont {Ambainis}\ \emph {et~al.}(1999)\citenamefont
  {Ambainis}, \citenamefont {Nayak}, \citenamefont {Ta-Shma},\ and\
  \citenamefont {Vazirani}}]{Ambainis1999}%
  \BibitemOpen
  \bibfield  {author} {\bibinfo {author} {\bibfnamefont {A.}~\bibnamefont
  {Ambainis}}, \bibinfo {author} {\bibfnamefont {A.}~\bibnamefont {Nayak}},
  \bibinfo {author} {\bibfnamefont {A.}~\bibnamefont {Ta-Shma}},\ and\ \bibinfo
  {author} {\bibfnamefont {U.}~\bibnamefont {Vazirani}},\ }\bibfield  {title}
  {\bibinfo {title} {Dense quantum coding and a lower bound for 1-way quantum
  automata},\ }in\ \href {https://doi.org/10.1145/301250.301347} {\emph
  {\bibinfo {booktitle} {Proceedings of the thirty-first annual ACM symposium
  on Theory of Computing}}},\ \bibinfo {series and number} {STOC '99}\
  (\bibinfo  {publisher} {Association for Computing Machinery},\ \bibinfo
  {address} {New York, NY, USA},\ \bibinfo {year} {1999})\ pp.\ \bibinfo
  {pages} {376--383}\BibitemShut {NoStop}%
\bibitem [{\citenamefont {Tavakoli}\ \emph {et~al.}(2021)\citenamefont
  {Tavakoli}, \citenamefont {Pauwels}, \citenamefont {Woodhead},\ and\
  \citenamefont {Pironio}}]{Tavakoli2021a}%
  \BibitemOpen
  \bibfield  {author} {\bibinfo {author} {\bibfnamefont {A.}~\bibnamefont
  {Tavakoli}}, \bibinfo {author} {\bibfnamefont {J.}~\bibnamefont {Pauwels}},
  \bibinfo {author} {\bibfnamefont {E.}~\bibnamefont {Woodhead}},\ and\
  \bibinfo {author} {\bibfnamefont {S.}~\bibnamefont {Pironio}},\ }\bibfield
  {title} {\bibinfo {title} {Correlations in entanglement-assisted
  prepare-and-measure scenarios},\ }\href
  {https://doi.org/10.1103/PRXQuantum.2.040357} {\bibfield  {journal} {\bibinfo
   {journal} {PRX Quantum}\ }\textbf {\bibinfo {volume} {2}},\ \bibinfo {pages}
  {040357} (\bibinfo {year} {2021})}\BibitemShut {NoStop}%
\bibitem [{\citenamefont {Pauwels}\ \emph
  {et~al.}(2022{\natexlab{b}})\citenamefont {Pauwels}, \citenamefont
  {Tavakoli}, \citenamefont {Woodhead},\ and\ \citenamefont
  {Pironio}}]{Pauwels2022a}%
  \BibitemOpen
  \bibfield  {author} {\bibinfo {author} {\bibfnamefont {J.}~\bibnamefont
  {Pauwels}}, \bibinfo {author} {\bibfnamefont {A.}~\bibnamefont {Tavakoli}},
  \bibinfo {author} {\bibfnamefont {E.}~\bibnamefont {Woodhead}},\ and\
  \bibinfo {author} {\bibfnamefont {S.}~\bibnamefont {Pironio}},\ }\bibfield
  {title} {\bibinfo {title} {Entanglement in prepare-and-measure scenarios:
  many questions, a few answers},\ }\href
  {https://doi.org/10.1088/1367-2630/ac724a} {\bibfield  {journal} {\bibinfo
  {journal} {New Journal of Physics}\ }\textbf {\bibinfo {volume} {24}},\
  \bibinfo {pages} {063015} (\bibinfo {year} {2022}{\natexlab{b}})}\BibitemShut
  {NoStop}%
\bibitem [{\citenamefont {Weinfurter}(1994)}]{Weinfurter1994}%
  \BibitemOpen
  \bibfield  {author} {\bibinfo {author} {\bibfnamefont {H.}~\bibnamefont
  {Weinfurter}},\ }\bibfield  {title} {\bibinfo {title} {Experimental
  bell-state analysis},\ }\href {https://doi.org/10.1209/0295-5075/25/8/001}
  {\bibfield  {journal} {\bibinfo  {journal} {Europhysics Letters}\ }\textbf
  {\bibinfo {volume} {25}},\ \bibinfo {pages} {559} (\bibinfo {year}
  {1994})}\BibitemShut {NoStop}%
\bibitem [{\citenamefont {Braunstein}\ and\ \citenamefont
  {Mann}(1995)}]{Braunstein1995}%
  \BibitemOpen
  \bibfield  {author} {\bibinfo {author} {\bibfnamefont {S.~L.}\ \bibnamefont
  {Braunstein}}\ and\ \bibinfo {author} {\bibfnamefont {A.}~\bibnamefont
  {Mann}},\ }\bibfield  {title} {\bibinfo {title} {Measurement of the bell
  operator and quantum teleportation},\ }\href
  {https://doi.org/10.1103/PhysRevA.51.R1727} {\bibfield  {journal} {\bibinfo
  {journal} {\pra}\ }\textbf {\bibinfo {volume} {51}},\ \bibinfo {pages}
  {R1727} (\bibinfo {year} {1995})}\BibitemShut {NoStop}%
\bibitem [{\citenamefont {Tavakoli}\ \emph {et~al.}(2015)\citenamefont
  {Tavakoli}, \citenamefont {Hameedi}, \citenamefont {Marques},\ and\
  \citenamefont {Bourennane}}]{Tavakoli2015}%
  \BibitemOpen
  \bibfield  {author} {\bibinfo {author} {\bibfnamefont {A.}~\bibnamefont
  {Tavakoli}}, \bibinfo {author} {\bibfnamefont {A.}~\bibnamefont {Hameedi}},
  \bibinfo {author} {\bibfnamefont {B.}~\bibnamefont {Marques}},\ and\ \bibinfo
  {author} {\bibfnamefont {M.}~\bibnamefont {Bourennane}},\ }\bibfield  {title}
  {\bibinfo {title} {Quantum random access codes using single $d$-level
  systems},\ }\href {https://doi.org/10.1103/PhysRevLett.114.170502} {\bibfield
   {journal} {\bibinfo  {journal} {\prl}\ }\textbf {\bibinfo {volume} {114}},\
  \bibinfo {pages} {170502} (\bibinfo {year} {2015})}\BibitemShut {NoStop}%
\bibitem [{\citenamefont {Navascu\'es}\ \emph {et~al.}(2015)\citenamefont
  {Navascu\'es}, \citenamefont {Feix}, \citenamefont {Ara\'ujo},\ and\
  \citenamefont {V\'ertesi}}]{Navascues2015a}%
  \BibitemOpen
  \bibfield  {author} {\bibinfo {author} {\bibfnamefont {M.}~\bibnamefont
  {Navascu\'es}}, \bibinfo {author} {\bibfnamefont {A.}~\bibnamefont {Feix}},
  \bibinfo {author} {\bibfnamefont {M.}~\bibnamefont {Ara\'ujo}},\ and\
  \bibinfo {author} {\bibfnamefont {T.}~\bibnamefont {V\'ertesi}},\ }\bibfield
  {title} {\bibinfo {title} {Characterizing finite-dimensional quantum
  behavior},\ }\href {https://doi.org/10.1103/PhysRevA.92.042117} {\bibfield
  {journal} {\bibinfo  {journal} {\pra}\ }\textbf {\bibinfo {volume} {92}},\
  \bibinfo {pages} {042117} (\bibinfo {year} {2015})}\BibitemShut {NoStop}%
\bibitem [{\citenamefont {Tavakoli}\ \emph {et~al.}(2020)\citenamefont
  {Tavakoli}, \citenamefont {Zambrini~Cruzeiro}, \citenamefont {Bohr~Brask},
  \citenamefont {Gisin},\ and\ \citenamefont {Brunner}}]{InfoCorrelations1}%
  \BibitemOpen
  \bibfield  {author} {\bibinfo {author} {\bibfnamefont {A.}~\bibnamefont
  {Tavakoli}}, \bibinfo {author} {\bibfnamefont {E.}~\bibnamefont
  {Zambrini~Cruzeiro}}, \bibinfo {author} {\bibfnamefont {J.}~\bibnamefont
  {Bohr~Brask}}, \bibinfo {author} {\bibfnamefont {N.}~\bibnamefont {Gisin}},\
  and\ \bibinfo {author} {\bibfnamefont {N.}~\bibnamefont {Brunner}},\
  }\bibfield  {title} {\bibinfo {title} {Informationally restricted quantum
  correlations},\ }\href {https://doi.org/10.22331/q-2020-09-24-332} {\bibfield
   {journal} {\bibinfo  {journal} {{Quantum}}\ }\textbf {\bibinfo {volume}
  {4}},\ \bibinfo {pages} {332} (\bibinfo {year} {2020})}\BibitemShut {NoStop}%
\bibitem [{\citenamefont {Tavakoli}\ \emph {et~al.}(2022)\citenamefont
  {Tavakoli}, \citenamefont {Zambrini~Cruzeiro}, \citenamefont {Woodhead},\
  and\ \citenamefont {Pironio}}]{InfoCorrelations2}%
  \BibitemOpen
  \bibfield  {author} {\bibinfo {author} {\bibfnamefont {A.}~\bibnamefont
  {Tavakoli}}, \bibinfo {author} {\bibfnamefont {E.}~\bibnamefont
  {Zambrini~Cruzeiro}}, \bibinfo {author} {\bibfnamefont {E.}~\bibnamefont
  {Woodhead}},\ and\ \bibinfo {author} {\bibfnamefont {S.}~\bibnamefont
  {Pironio}},\ }\bibfield  {title} {\bibinfo {title} {Informationally
  restricted correlations: a general framework for classical and quantum
  systems},\ }\href {https://doi.org/10.22331/q-2022-01-05-620} {\bibfield
  {journal} {\bibinfo  {journal} {{Quantum}}\ }\textbf {\bibinfo {volume}
  {6}},\ \bibinfo {pages} {620} (\bibinfo {year} {2022})},\ \Eprint
  {https://arxiv.org/abs/2007.16145} {arXiv:2007.16145 [quant-ph]} \BibitemShut
  {NoStop}%
\bibitem [{\citenamefont {Pironio}\ \emph {et~al.}(2010)\citenamefont
  {Pironio}, \citenamefont {Ac{\'\i}n}, \citenamefont {Massar}, \citenamefont
  {de~La~Giroday}, \citenamefont {Matsukevich}, \citenamefont {Maunz},
  \citenamefont {Olmschenk}, \citenamefont {Hayes}, \citenamefont {Luo},
  \citenamefont {Manning} \emph {et~al.}}]{Pironio2010}%
  \BibitemOpen
  \bibfield  {author} {\bibinfo {author} {\bibfnamefont {S.}~\bibnamefont
  {Pironio}}, \bibinfo {author} {\bibfnamefont {A.}~\bibnamefont {Ac{\'\i}n}},
  \bibinfo {author} {\bibfnamefont {S.}~\bibnamefont {Massar}}, \bibinfo
  {author} {\bibfnamefont {A.~B.}\ \bibnamefont {de~La~Giroday}}, \bibinfo
  {author} {\bibfnamefont {D.~N.}\ \bibnamefont {Matsukevich}}, \bibinfo
  {author} {\bibfnamefont {P.}~\bibnamefont {Maunz}}, \bibinfo {author}
  {\bibfnamefont {S.}~\bibnamefont {Olmschenk}}, \bibinfo {author}
  {\bibfnamefont {D.}~\bibnamefont {Hayes}}, \bibinfo {author} {\bibfnamefont
  {L.}~\bibnamefont {Luo}}, \bibinfo {author} {\bibfnamefont {T.~A.}\
  \bibnamefont {Manning}}, \emph {et~al.},\ }\bibfield  {title} {\bibinfo
  {title} {Random numbers certified by bell’s theorem},\ }\href
  {https://doi.org/10.1038/nature09008} {\bibfield  {journal} {\bibinfo
  {journal} {Nature}\ }\textbf {\bibinfo {volume} {464}},\ \bibinfo {pages}
  {1021} (\bibinfo {year} {2010})}\BibitemShut {NoStop}%
\bibitem [{\citenamefont {Gill}(2002)}]{Gill2002}%
  \BibitemOpen
  \bibfield  {author} {\bibinfo {author} {\bibfnamefont {R.~D.}\ \bibnamefont
  {Gill}},\ }\bibfield  {title} {\bibinfo {title} {Time, finite statistics, and
  bell's fifth position},\ }\href {https://doi.org/10.1.1.6.5788} {\bibfield
  {journal} {\bibinfo  {journal} {Foundations of Probability and Physics}\
  }\textbf {\bibinfo {volume} {5}},\ \bibinfo {pages} {179} (\bibinfo {year}
  {2002})}\BibitemShut {NoStop}%
\bibitem [{\citenamefont {Ambainis}\ \emph {et~al.}(2015)\citenamefont
  {Ambainis}, \citenamefont {Kravchenko},\ and\ \citenamefont
  {Rai}}]{Ambainis2015}%
  \BibitemOpen
  \bibfield  {author} {\bibinfo {author} {\bibfnamefont {A.}~\bibnamefont
  {Ambainis}}, \bibinfo {author} {\bibfnamefont {D.}~\bibnamefont
  {Kravchenko}},\ and\ \bibinfo {author} {\bibfnamefont {A.}~\bibnamefont
  {Rai}},\ }\href@noop {} {\bibinfo {title} {{O}ptimal {C}lassical {R}andom
  {A}ccess {C}odes {U}sing {S}ingle d-level {S}ystems}} (\bibinfo {year}
  {2015}),\ \bibinfo {note} {arXiv:1510.03045v1},\ \Eprint
  {https://arxiv.org/abs/1510.03045} {1510.03045} \BibitemShut {NoStop}%
\bibitem [{\citenamefont {Anwer}\ \emph {et~al.}(2020)\citenamefont {Anwer},
  \citenamefont {Muhammad}, \citenamefont {Cherifi}, \citenamefont {Miklin},
  \citenamefont {Tavakoli},\ and\ \citenamefont {Bourennane}}]{Anwer2020}%
  \BibitemOpen
  \bibfield  {author} {\bibinfo {author} {\bibfnamefont {H.}~\bibnamefont
  {Anwer}}, \bibinfo {author} {\bibfnamefont {S.}~\bibnamefont {Muhammad}},
  \bibinfo {author} {\bibfnamefont {W.}~\bibnamefont {Cherifi}}, \bibinfo
  {author} {\bibfnamefont {N.}~\bibnamefont {Miklin}}, \bibinfo {author}
  {\bibfnamefont {A.}~\bibnamefont {Tavakoli}},\ and\ \bibinfo {author}
  {\bibfnamefont {M.}~\bibnamefont {Bourennane}},\ }\bibfield  {title}
  {\bibinfo {title} {Experimental characterization of unsharp qubit observables
  and sequential measurement incompatibility via quantum random access codes},\
  }\href {https://doi.org/10.1103/PhysRevLett.125.080403} {\bibfield  {journal}
  {\bibinfo  {journal} {\prl}\ }\textbf {\bibinfo {volume} {125}},\ \bibinfo
  {pages} {080403} (\bibinfo {year} {2020})}\BibitemShut {NoStop}%
\bibitem [{\citenamefont {Hong}\ \emph {et~al.}(1987)\citenamefont {Hong},
  \citenamefont {Ou},\ and\ \citenamefont {Mandel}}]{Hong1987}%
  \BibitemOpen
  \bibfield  {author} {\bibinfo {author} {\bibfnamefont {C.~K.}\ \bibnamefont
  {Hong}}, \bibinfo {author} {\bibfnamefont {Z.~Y.}\ \bibnamefont {Ou}},\ and\
  \bibinfo {author} {\bibfnamefont {L.}~\bibnamefont {Mandel}},\ }\bibfield
  {title} {\bibinfo {title} {Measurement of subpicosecond time intervals
  between two photons by interference},\ }\href
  {https://doi.org/10.1103/PhysRevLett.59.2044} {\bibfield  {journal} {\bibinfo
   {journal} {Phys. Rev. Lett.}\ }\textbf {\bibinfo {volume} {59}},\ \bibinfo
  {pages} {2044} (\bibinfo {year} {1987})}\BibitemShut {NoStop}%
\end{thebibliography}%

\appendix
\onecolumngrid

\section*{Supplementary material}

Here we provide additional details on the experimental setup including details about the parties' unitary operations and how they are implemented, experimental error estimation and two-fold Hong-Ou-Mandel dip visibility.

\subsection{Random Access Code - $\mathcal{R}$}
Consider a Random Access Code based on multi-valued inputs \cite{Tavakoli2015}. The sender has a total of 16 inputs, expressed as two pieces of data $x_1,x_2\in\{1,2,3,4\}$. The receiver randomly selects the data of interest, either choosing $y=1$ for $x_1$ or $y=2$ for $x_2$. The aim is for the sender to encode $(x_1,x_2)$ into a message such that the receiver can recover $x_y$ with high average probability. The average success rate is given by $\mathcal{R}=\frac{1}{32}\sum_{x_1,x_2,y}p(b=x_y|x_1,x_2,y)$.

%Let the parties share the EPR state $\ket{\phi^+}=\frac{1}{\sqrt{2}}\left[\ket{00}+\ket{11}\right]$. To encode her message, the sender applies a unitary $U_{x_1x_2}=\vec{n}_{x_1x_2}\cdot \vec{\sigma}$ to her share of the EPR state, where we may write the Bloch vectors as $\vec{n}_{x_1x_2}=\left[\cos\phi_{x_1x_2} \sin \varphi_{x_1x_2},\sin\phi_{x_1x_2}\sin\varphi_{x_1x_2},\cos\varphi_{x_1x_2}\right]$ \jef{why introduce this notation if we never use it?}. The measurements of the receiver corresponds to the positive operator-valued measure $\{E_{b|y}\}_{b,y}$, where $E_{b|y}\geq 0$ and $\sum_b E_{b|y}=\openone$. In a quantum model, the success rate becomes
%\begin{equation}\label{rac}
%\mathcal{R}=\frac{1}{32}\sum_{x_1,x_2,y}\Tr\left[\left(U_{x_1x_2}\otimes \openone\right)\ketbra{\phi^+}{\phi^+}\left(U_{x_1x_2}^\dagger\otimes \openone\right)E_{x_y|y} \right].
%\end{equation}
In the main text, we gave an explicit optimal strategy that achieves $\mathcal{R}=\frac{3}{4}$ in which the two measurements are mutually unbiased bases of maximally entangled states. It is  interesting to note that in this quantum protocol, all winning probabilities are equal, i.e.~$\forall (x_1,x_2,y)$ we have $p(b=x_y|x_1,x_2,y)=\frac{3}{4}$. Hence, the worst-case probability of recovering the data of interest is equal to the average probability $\mathcal{R}$.

If the EPR state is subject to isotropic noise, i.e.~the effective state is $v\ketbra{\phi^+}{\phi^+}+\frac{1-v}{4}\openone$, then the optimal strategy returns an advantage over protocols based on two bits of communication whenever $v>3/4$. This is found immediately from solving $\frac{3}{4}v+(1-v)\frac{1}{4}=\frac{5}{8}$. The reason is that with probability $v$ one plays the optimal strategy, achieving $\mathcal{R}=\frac{3}{4}$, while with probability $1-v$ one randomly guesses the output, achieving $\mathcal{R}=\frac{1}{4}$, which has to equate to the optimal success rate based on 2 bits, which is  $\mathcal{R}=\frac{5}{8}$ \cite{Ambainis2015}.

\begin{figure}[h!]
	\includegraphics{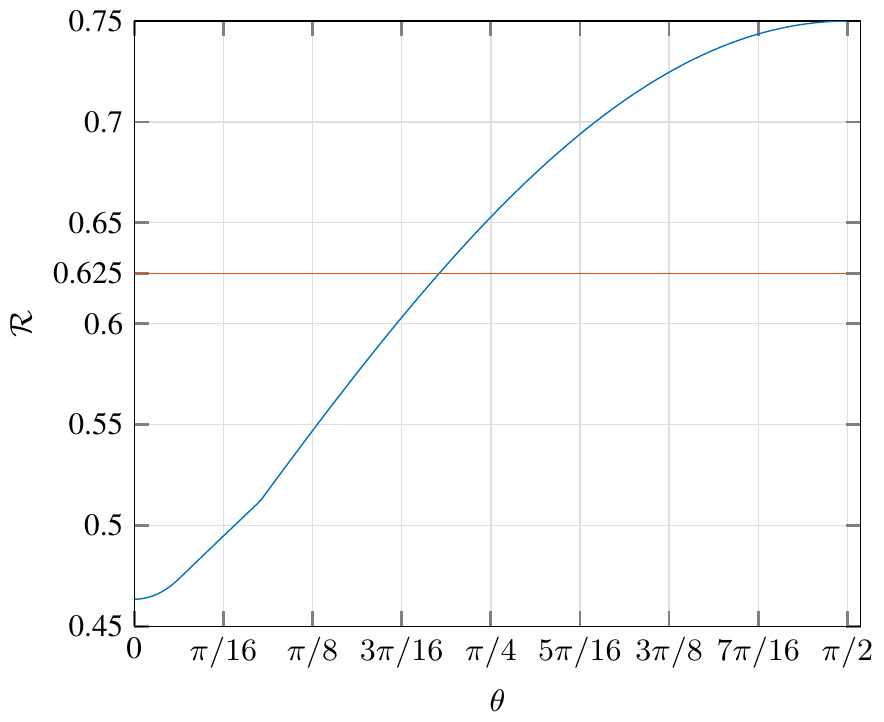}
	\caption{Numerically obtained success rate in the Random Access Code versus the degree of entanglement in the shared state. The red horizontal line represents the best success rate based on 2 bits of communication. The success rate decreases as the entanglement weakens, because this reduces the dense coding capability  of the channel. Above $\theta\gtrsim 0.6720$ the dense coding capability is still large enough to allow a quantum strategy to beat the success rate of 2 bit protocols.}\label{FigTradeoff}
\end{figure}

The use of a maximally entangled qubit pair is essential in order to convey two bits with a single qubit message. To further elucidate the relationship between information capabilities and correlations created, we instead assign a pure partially entangled state $\ket{\psi_\theta}=\cos\frac{\theta}{2} \ket{00}+\sin\frac{\theta}{2}\ket{11}$, for $\theta\in[0,\frac{\pi}{2}]$. One expects the ability to convey information to grow with $\theta$, from sending one bit when $\theta=0$ to sending two bits when $\theta=\frac{\pi}{2}$. A natural way to quantify the information capability of such a partially entangled state is to consider the success rate with which the state can be used in dense coding. In other words, the largest probability of recovering a four-valued message $x$ by applying a unitary transformation $U_x$ to one share of the state and then extracting the information via a quantum measurement $\{M_b\}$. We therefore write
\begin{equation}
\mathcal{D}(\psi_\theta)=\max \frac{1}{4}\sum_{x=1}^4 p(b=x|x,\psi_\theta)= \max_{\{U_x\},\{M_b\}}\frac{1}{4}\sum_{x=1}^4 \Tr\left[\left(U_{x}\otimes \openone\right)\ketbra{\psi_\theta}{\psi_\theta}\left(U_{x}^\dagger\otimes \openone\right)M_x\right].
\end{equation}
Following dense coding, we consider that the extraction measurement is a Bell basis measurement up to a local unitary, i.e.~$M_b=\ketbra{B_b}{B_b}$ with $\ket{B_b}=V\otimes \openone \ket{\phi_b}$, where the four states $\ket{\phi_b}$ are the Bell states $\ket{\phi^+},\ket{\phi^-},\ket{\psi^+}$ and $\ket{\psi^-}$. This simplifies the dense coding ability to
\begin{equation}
\mathcal{D}(\psi_\theta)=\max_{\{W_x\}}\frac{1}{4}\sum_{x=1}^4 |\bracket{\phi_x}{\left(W_{x}\otimes \openone\right)}{\psi_\theta}|^2,
\end{equation}
where $W_x=V^\dagger U_x$. One can then see that the optimal unitaries are the same as in dense coding, namely $W_1=\openone$, $W_2=\sigma_Z$, $W_3=\sigma_X$ and $W_4=\sigma_Y$. This gives $\mathcal{D}(\psi_\theta)=\frac{1+\sin\theta}{2}$. As expected, we see that the dense coding ability increases from $50\%$ ($\theta=0$) to $100\%$ ($\theta=\frac{\pi}{2}$). We consider the critical degree of entanglement, i.e.~the smallest value of $\theta$, such that we can leverage a qubit message and a shared state $\ket{\psi_\theta}$ to beat the limit of a 2 bit strategy in the Random Access Code. We have numerically investigated this and the results are shown in Figure~\ref{FigTradeoff}. As expected, we find that the success rate decreases as the dense coding capability decreases. Nevertheless, we are able to outperform the limit $\mathcal{R}=\frac{5}{8}$ whenever $\theta\gtrsim 0.6720$. This corresponds to $\mathcal{D}\approx 81\%$.% and to a one-shot accessible information \cite{Konig} of $I(X:B)=H_\text{min}(X)-H_\text{min}(X|B)=-\log_2\left(\frac{1}{4}\right)+\log_2\left(\mathcal{D}\right)\approx 1.69$ bits. 
Thus, quantum resources of a significantly smaller capacity still outperform two bits of communication in the Random Access Code. 

\subsection{Unitary operations in the experiment corresponding to figure of merit $\mathcal{S}$}

Depending on her setting $x$, the sender applies the unitary transformations $U^{\rm S}_x$, given by
\begin{align}\nonumber
&U^\text{S}_1=\openone, \quad U^\text{S}_2=\frac{-\sigma_Z\sqrt{3}-\sigma_X}{2}, \quad U^\text{S}_3=\frac{\sigma_X\sqrt{3}-\sigma_Z}{2},\\
&\quad \quad \quad U^\text{S}_4=\frac{\openone-i\sigma_Y\sqrt{3}}{2},\quad U^\text{S}_5=\frac{\openone+i\sigma_Y\sqrt{3}}{2} \, .
\end{align} 

The receiver, in turn depending on his setting $y$, applies the unitary transformation $U^\text{R}_y$, given by 
\begin{align}\nonumber
&U^\text{R}_1=\openone, \quad U^\text{R}_2=\frac{\nu_+\openone+i\nu_-\sigma_Y}{2\sqrt{2}}, \quad U^\text{R}_3=\frac{\nu_+\openone-i\nu_-\sigma_Y}{2\sqrt{2}},\\
& \qquad \quad U^\text{R}_4= U^\text{S}_2, \quad U^\text{R}_5=U^\text{S}_3,\quad U^\text{R}_6=\frac{\openone-i\sigma_Y}{\sqrt{2}} \, ,
\end{align} 
where $\nu_\pm=\sqrt{3}\pm1$. These unitary operations are realised in the experiment by rotating a half wave plate with rotation angle $\theta$ combined with a phase shift $\phi$,

\begin{equation}
U^\text{\rm K}_z=   \begin{pmatrix} \cos(2\theta_{\rm K}^z) & \sin(2\theta_{\rm K}^z)\\ e^{i\phi_{\rm K}^z } \sin(2\theta_{\rm K}^z) & - e^{i \phi_{\rm K}^z} \cos(2\theta_{\rm K}^z) \end{pmatrix} \,,
\end{equation} 
where $z=x$ for $\rm K = S$ and $z=y$ for $\rm K = R$.

The rotation angles $\theta_{\rm K}^z$ and phases $\phi_{\rm K}^z$ corresponding to each setting for the sender and receiver are listed in Table~\ref{tab:1} and Table~\ref{tab:2} respectively.

\begin{table}[h!]
	\begin{ruledtabular}
		\begin{tabular}{lll}$x $&$\phi_{\rm S}^x$ &$\theta_{\rm S}^x$\\
			\midrule
			1&$\pi$&$0$\\
			2&$0$&15\\
			3&$0$&-30\\
			4&$\pi$& -30\\
			5 & $\pi$ & 30\\
		\end{tabular}
	\end{ruledtabular}
	\caption{\label{tab:1} Half wave plate rotation angles $\theta_{\rm S}^x$ and phase shifts $\phi_{\rm S}^x$ applied by the sender for different settings $x$. }
\end{table}

\begin{table}[h!]
	\begin{ruledtabular}
		\begin{tabular}{lll}$y$&$\phi_{\rm R}^y$ &$\theta_{\rm R}^y$\\
			\midrule
			1&$\pi$&0\\
			2&$\pi$&7.5\\
			3&$\pi$&-7.5\\
			4&$0$&15\\
			5&$0$&-30\\
			6&$\pi$&-22.5\\
		\end{tabular}
	\end{ruledtabular}
	\caption{\label{tab:2}Half wave plate rotation angles $\theta_{\rm R}^y$ and phase shifts $\phi_{\rm R}^y$ applied by the receiver for different settings $y$.}
\end{table}

\subsection{Experimental results for $\mathcal{S}$}

Table~\ref{tab:3} lists our experimental results alongside the theoretical probabilities for each combination of settings $(x,y)$ associated to a non-zero payoff in the communication task ($c_{xy}\neq 0$). We also list the estimated errors, discussed below. 

\begin{table}[h]
	
	\begin{ruledtabular}
		\begin{tabular}{llll}$c_{xy}$&$p(b|x,y)$  &$p(b|x,y)$ &Errors\\
			&Theory &Experiment&\\
			\midrule
			$c_{11}$&1&0.9725&0.0003\\
			$c_{21}$&0&0.0250&0.002\\
			$c_{31}$&0&0.0100&0.003\\
			$c_{12}$&$(2 + \sqrt{3})/4$&0.9087&0.0006\\
			$c_{42}$&$(2 - \sqrt{3})/4$&0.0610&0.001\\
			$c_{13}$&$(2 + \sqrt{3})/4$&0.9198&0.0006\\
			$c_{53}$&$(2 - \sqrt{3})/4$&0.0793&0.0009\\
			$c_{24}$&1&0.9688&0.0003\\
			$c_{34}$&0&0.0020&0.005\\
			$c_{44}$&0&0.0050&0.004\\
			$c_{54}$&0&0.0120&0.002\\
			$c_{35}$&1&0.9828&0.0003\\
			$c_{45}$&0&0.0240&0.002\\
			$c_{55}$&0&0.0070&0.003\\
			$c_{46}$&$(2 + \sqrt{3})/4$&0.9068&0.0006\\
			$c_{56}$&$(2 - \sqrt{3})/4$&0.0560&0.001\\
			
		\end{tabular}
	\end{ruledtabular}
	\caption{\label{tab:3}Experimental results for $\mathcal{S}$. }
\end{table}

\subsection{Unitary operations in the experiment corresponding to figure of merit $\mathcal{T}$}

Depending on her setting $x_{1},x_{2},x_{3}$, the sender applies the unitary transformations $U_x$, given by:

\begin{equation}
U_x=(-1)^{x_1}
\begin{pmatrix}
-\alpha_{x_1}\mu_{x_2x_3} & (-1)^{x_2+x_3}\alpha_{\bar{x}_1}\mu_{x_2x_3}\\
(-1)^{x_2+x_3}\sqrt{2}\alpha_{\bar{x}_1} & \sqrt{2}\alpha_{x_1},
\end{pmatrix} \, ,
\end{equation}
where $\mu_{x_2x_3}=(-1)^{x_2}+i(-1)^{x_3}$, $\alpha_s=\frac{1}{2}\sqrt{1+(-1)^s\sqrt{2/3}}$ and the bar-sign denotes bit-flip. These unitary operations are realised in the experiment by two half wave plates and two quarter wave plates combined with a phase shift. The receiver makes the following measurements:
\begin{center}
	\begin{equation}
	E_1=\sigma_Z\otimes \sigma_Z \, , \quad
	E_2=\frac{1}{2}\sigma_Y\otimes (\sqrt{3}\sigma_Y+\sigma_Z) \,, \quad
	E_3=\frac{1}{2} \sigma_X \otimes (\sqrt{3}\sigma_Y-\sigma_Z) \, .
	\end{equation}
\end{center}
These measurements are realised in the experiment by a half wave plate and a quarter wave plate on the photon receiving the unitary and with a half wave plate with two quater on the other photon. The rotation angles and phases corresponding to each setting for the sender and receiver are listed in Table~\ref{tab:4} and Table~\ref{tab:5} respectively. Wave plates are named in the order of arrival for the photon, as shown in the figure in the main text.

\begin{table}[h]
	\begin{ruledtabular}
		\begin{tabular}{llllll} $U_{x_{1}x_{2}x_{3}}$ &phase&$H_{1}$&$Q_{1}$&$H_{2}$&$Q_{2}$ \\
			\midrule
			$U_{000}$&$\pi$&-8.816&45&33.75&45\\
			$U_{001}$&0&-8.816&45&-78.75&45\\
			$U_{010}$&0&-8.816&45&-33.75&45\\
			$U_{011}$&$\pi$&-8.816&45&78.75&45\\
			$U_{100}$&$\pi$&53.816&45&33.75&45\\
			$U_{101}$&0&53.816&45&-78.75&45\\
			$U_{110}$&0&53.816&45&-33.75&45\\
			$U_{111}$&$\pi$&53.816&45&78.75&45\\
		\end{tabular}
	\end{ruledtabular}
	\caption{\label{tab:4} Half and quarter wave plate rotation angles and phase shifts applied by the sender for different settings $U_{x_{1}x_{2}x_{3}}$. }
\end{table}

\begin{table}[h]
	\begin{ruledtabular}
		\begin{tabular}{llllll}&Mode1&Mode1&Mode2&Mode2&Mode2\\
			$y$&$H_{1}$&$Q_{1}$&$Q_{1}$&$H_{1}$&$Q_{2}$\\
			\midrule
			1&0&0&0&0&0\\
			2&0&-45&0& 15&0\\
			3&22.5&0&0&30&0\\
		\end{tabular}
	\end{ruledtabular}
	\caption{\label{tab:5}Half and quarter wave plate rotation angles applied by the receiver on each arms for different settings $y$.}
\end{table}

\subsection{Experimental results for $\mathcal{T}$}

Table~\ref{tab:6} lists our experimental results for each combination of unitary rotation $U_{x_{1}x_{2}x_{3}}$ and measurent $E_{y}$. We also list the estimated errors, as for the other experiment.

\begin{table}[h]
	\begin{ruledtabular}
		\begin{tabular}{lllll} $U_{x_{1}x_{2}x_{3}}$ & Measurement & $p(b=x_{y}|x,y)$  & $p(b=x_{y}|x,y)$ & error \\
			& &Theory &Experiment&\\
			\midrule
			$U_{000}$ &$E_{1}$&0.9082&0.9139& 0.01\\
			$U_{000}$ &$E_{2}$&0.9082&0.864& 0.01\\
			$U_{000}$ &$E_{3}$&0.9082&0.9062& 0.01\\
			$U_{001}$ &$E_{1}$&0.9082&0.9535& 0.01\\
			$U_{001}$ &$E_{2}$&0.9082&0.8968& 0.01\\
			$U_{001}$ &$E_{3}$&0.9082&0.8532& 0.01\\
			$U_{010}$ &$E_{1}$&0.9082&0.9814& 0.01\\
			$U_{010}$ &$E_{2}$&0.9082&0.8458& 0.01\\
			$U_{010}$ &$E_{3}$&0.9082&0.8808& 0.01\\
			$U_{011}$ &$E_{1}$&0.9082&0.9087& 0.01\\
			$U_{011}$ &$E_{2}$&0.9082&0.8753& 0.01\\
			$U_{011}$ &$E_{3}$&0.9082&0.9081& 0.01\\
			$U_{100}$ &$E_{1}$&0.9082&0.8979& 0.01\\
			$U_{100}$ &$E_{2}$&0.9082&0.9057& 0.01\\
			$U_{100}$ &$E_{3}$&0.9082&0.9044& 0.01\\
			$U_{101}$ &$E_{1}$&0.9082&0.8787& 0.01\\
			$U_{101}$ &$E_{2}$&0.9082&0.8746& 0.01\\
			$U_{101}$ &$E_{3}$&0.9082&0.9356& 0.01\\
			$U_{110}$ &$E_{1}$&0.9082&0.8247& 0.01\\
			$U_{110}$ &$E_{2}$&0.9082&0.9136& 0.01\\
			$U_{110}$ &$E_{3}$&0.9082&0.9266& 0.01\\
			$U_{111}$ &$E_{1}$&0.9082&0.9113& 0.01\\
			$U_{111}$ &$E_{2}$&0.9082&0.905& 0.01\\
			$U_{111}$ &$E_{3}$&0.9082&0.9044& 0.01\\
			\midrule
			$\mathcal{T}=\frac{1}{24}\sum_{x,y}p(b=x_y|x,y)$ &  &0.9082 &0.8988&0.003\\
		\end{tabular}
	\end{ruledtabular}
	\caption{\label{tab:6} Experimental results for $\mathcal{T}$.}
\end{table}

\subsection{Error estimation} Following \cite{Anwer2020} we consider error originating from the measurement side only. To reduce experimental errors in the measurements, we used computer controlled high precision motorised rotation stages to set the orientation of wave-plates with repeatability precision  $0.02 ^{\circ}$ for the first experiment and $0.025 ^{\circ}$ for the second experiment. The use of different settings $(x,y)$ induces a systematic error, which we estimate using Monte Carlo simulation. We assume that the wave-plates setting error is normally distributed with a standard deviation of $0.02^{\circ}$ for the first experiment and $0.025 ^{\circ}$ for the second experiment. This together with the Poissonian error in photon counting statistics comprise the final error reported here. Due to inefficiency in the single photon detectors, the photons are detected  randomly  and  their counting  is Poissonian. To decrease Poissonian counting error, we  have chosen a measurement time  of  two hours  for every setting and collected  about 18 Million events. To guarantee that both parties receive single qubits, we worked at a low rate ($\approx 2500$ pairs per sec)  to suppress higher order coincidence to almost $0.9$ per sec.  

\subsection{Two-fold Hong-Ou-Mandel dip visibility}

\begin{figure}[h!]
	\centering
	\includegraphics[scale=0.8]{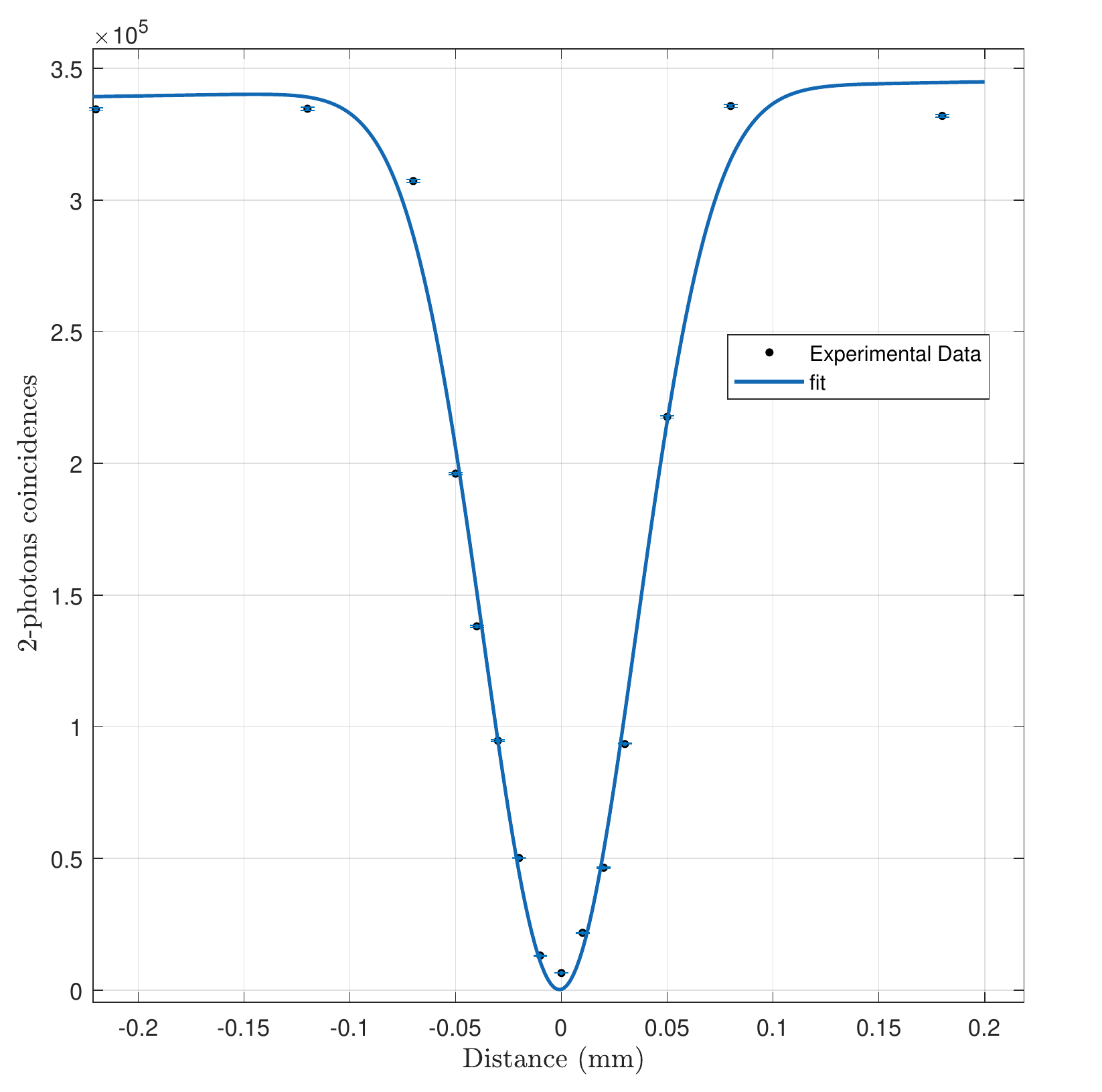}
	\caption{Two-fold Hong-Ou-Mandel dip. The plot displays the  two-fold photon counting coincidence  versus  the delay (the path  difference  between the two arms). The error bars  indicate  the  Poissonian photon counting error statistics. The  data is  fitted  with  Gaussian function.}\label{Fig1supp}
\end{figure}

Bell state measurements are implemented through  two-photon interference, using PBS and HWP plates set at 22.5°. The photons are detected by Si avalanche photodiodes and the coincidences are registered with an eight channels multifold coincidence counting unit. 
This Bell analyser consists of coherent interference at a polarisation beam splitter. To obtain indistinguishability of the photons, due to their arrival times, we adjusted the path length of one of the photons by using a delay line \cite{Hong1987}. In Figure~\ref{Fig1supp}, the coincidences between the detectors versus the delay path is shown.The zero delay corresponds to a maximal overlap (maximum indistinguishability). The interfering  photons  will bunch (they  will exit  only  in one output  arm of the PBS) causing the  coincidence to vanish.  The obtained visibility of the two-fold Hong-Ou-Mandel dip is $0.961 \pm 0.002$.

\end{document}